\documentstyle[lprocl,11pt,epsfig]{article}

\def\be{\begin{equation}}
\def\ee{\end{equation}}
\def\bea{\begin{eqnarray}}
\def\eea{\end{eqnarray}}

\begin{document}
\title{Statics and Dynamics of Disordered Elastic Systems}
\author{Thierry Giamarchi}
\address{Laboratoire de Physique des Solides\cite{junk}, UPS B\^at 510,
91405 Orsay, France, Email: giam@lps.u-psud.fr}
\author{Pierre Le Doussal}
\address{CNRS-LPTENS, 24 Rue Lhomond, 75230 Paris Cedex 05
France, Email: ledou@physique.ens.fr}
\maketitle\abstracts{
We examine here various aspects of the statics and dynamics of
disordered elastic systems such as manifolds and periodic systems.
Although these objects look very similar and indeed share some
underlying physics, periodic systems constitute a class of their own
with markedly different properties. We focus on such systems, review
the methods allowing to treat them, emphasize the shift of viewpoint
compared to the physics of manifolds and discuss their physics in detail.
As for the statics, periodicity helps
the system to retain a quasi-translational order and to be stable
with respect to the
proliferation of free topological defects such as dislocations. A
disordered periodic system thus leads to a glass phase with Bragg
peaks: the Bragg glass. On the other hand, for driven lattices,
transverse periodicity allows the system to retain its glassy nature,
leading to a moving glass phase. The existence of these two phases
has important theoretical and experimental consequences, in
particular for vortex physics in superconductors, the physical
system which is mainly focused here.
}

\section{Introduction}

The statics and dynamics of lines in random media is a long standing
problem
in the physics of disordered systems. It is one of the remarkable, and
experimentally relevant examples of glassy systems with strong analogies
but also marked differences with spin glasses.
Before the advent of the vortex problem in high Tc superconductors, and the strong
motivation to solve it that it entailed, the physics of manifold was
viewed as a sort of toy model for more ``complicated'' and ``important''
systems such as spin glasses.  Indeed, as a purely theoretical problem,
disordered models such as the random
field XY model, were thought to be at least qualitatively well
understood. The discovery of high-Tc superconductors and the new
experimental realizations it provided for these systems, shook this
belief and led to many new questions. The field
was then able to progress at a rapid pace due to the remarkable
interplay between the theory and experiments. It has
proven to possess a remarkable amount of novel and complex physics,
unexpected some ten years ago.

Since the focus of the community was on lines and manifolds, the first
theoretical papers on the vortex systems mainly borrowed from
this physics. Although this led to spectacular results in the field
of superconductors, it also fell short of the mark in some respects,
since it misses a good part of the novel physics arising for
such {\it periodic} objects and {\it entirely} due to the periodic
nature of the system. Indeed it was then realized that, both for the
statics and the dynamics, periodicity was a crucial ingredient that
needed to be treated carefully. Once taken into account, periodicity
led not only to new concepts
and methods, but also to a radically new physical image, that
replaced the previous common beliefs, based on manifold
physics. Fortunately, these new theoretical ideas coincided with an
experimental maturation of the field: experiments
became accurate, the main spurious effects were fairly well understood,
so that experimental results are now firm and unavoidable.
They put stringent constraints on whatever theoretical
interpretation can be proposed, well above the level that can be
satisfied by vague and very often self-contradictory ``handwaving''
arguments, and provide a very strong stimulus for accurate (and
correct) theories.

Since a number of reviews already exist in this field,
whether for the
statistical mechanics of the manifold point of view
\cite{kardar_review_lines}, or centered on the
physics of vortex systems
\cite{blatter_vortex_review,brandt_review_superconductors},
this paper specially develops
the new results, concepts and methods, derived in the last
four years, following from the treatment of periodicity.
Since the main actor in this progress are the vortex systems, we first
start in section~\ref{vorsec} by a very basic introduction to the
physics and important questions for such vortex lattices. We also
briefly review the various theoretical
approaches put forward to tackle this problem.
We then focus in section~\ref{statsec} on the general problem of
the statics of a periodic structure. Although such a study applies
to the vortex
lattice it has many other physical applications that we also briefly
discuss. A section is devoted to the interesting case of $d=2$.
section~\ref{dynsec} is devoted to the dynamical effects. Here
again periodicity plays a crucial role. After having reviewed the basic
methods and results for the dynamics of simple manifolds we concentrate
on these novel effects. Finally some
conclusions and open questions can be found in section~\ref{concsec}.

\section{Vortex physics}  \label{vorsec}

\subsection{Experimental questions}

The conventional mean field phase diagram
\cite{tinkham_book_superconductors} of type II superconductor
consists of a Meissner phase ($H<H_{c1}(T)$) and a mixed phase
($H_{c1}(T)< H < H_{c2}(T)$). In the
mixed phase the magnetic induction $B$ can
penetrate the bulk of the superconductor in the form of vortex lines each
carrying a quantum of
magnetic flux $\Phi_0 = h c/ 2 e$ and aligned along the external field
$H$. A vortex consists of a normal region of radius
$\xi_0$ where the superconducting order parameter $\Psi$ vanishes,
surrounded by a region of size $\lambda$ where supercurrents
screen the external field. By minimization of the
Landau-Ginzburg functional, Abrikosov predicted
\cite{abrikosov_vortex_first} that these vortex lines form a regular
lattice
(triangular in standard systems), later observed.
This lattice can also be described as a
standard lattice with elastic coefficients
\cite{degennes_matricon_supraelastic,blatter_vortex_review}. From flux
quantization, the lattice spacing is
simply related to the field
$a \sim \sqrt{\Phi_0/B}$. In high $T_c$ materials it can
vary over several orders of magnitudes
typically from $a \sim \lambda \sim 0.5 \mu m$
near $H_{c1}$ to $a \sim \xi_0 \sim 10A$ near $H_{c2}$
(outside the critical regions).

Even for a pure system (i.e. without disorder),
real life is more complicated and, as can be expected
from a lattice the Abrikosov
vortex lattice melts on a line $H_m(T)$ below $H_{c2}(T)$, as
predicted a long time ago \cite{eilenberger_vortex_fusion}. However
it is only with the advent of high-Tc superconductors, where
$H_m(T)$ is expected to lie well below $H_{c2}(T)$
due to higher anisotropy and temperatures, that
such effects were studied in details
\cite{blatter_vortex_review,nelson_fusion_vortex,houghton_fusion_vortex}.
The high temperature phase is a vortex liquid, which
in some regimes can be thought of as a collection of fluctuating entangled
lines \cite{nelson_fusion_vortex,nelson_seung_long,nelson_ledoussal_liquid}. For the pure
system, it is now reasonably established, mainly
through numerical simulations \cite{hetzel_firstorder_melting,sasik_stroud_melting} that
this transition occurs and is first order. The detailed
theory of this transition is difficult and still controversial
since it should describe both
the formation of the Abrikosov lattice and
the fluctuations of the superconducting order parameter
\cite{brezin_nelson_hc2,moore_frg_melting}.

An additional but necessary complication when dealing with
any superconducting phase is quenched disorder, as can be
seen from dynamical considerations. Indeed,
in the presence of a current $J$, each vortex line
segment $dl$ is submitted to a Lorentz force
$F= (\Phi_0/c) J \wedge dl$. In the absence of any
source of pinning the vortices will start moving
in the direction perpendicular to the current. This
motion will in turn generate an e.m.f. (a voltage drop)
along the same direction than the current and lead to dissipation:
the material will then be a rather poor conductor whose
resistivity will be $\rho_{\rm ff} = \rho_N (B/H_{\rm c2})$
(the so called flux flow resistivity) where $\rho_N$ is
the normal state resistivity. Disorder, on the other hand
provides preferential regions for the vortex cores to
sit in and thus prevents the vortex lattice from sliding freely
and dissipating. Thus, somewhat paradoxically, some amount
of quenched disorder is crucial to make the material a superconductor
(it is a fine balance since too much disorder is detrimental again
by destroying superconductivity altogether).
At $T=0$, the pinning by disorder leads to the existence
of a critical force $F_c$ (corresponding to a critical current density $J_c$), below which
vortices stay still, the average velocity is zero and no dissipation
occurs. Disorder already exists in any real material either
in the form of microscopic defects such as oxygen vacancies ({\it uncorrelated disorder})
or more macroscopic defects such as twin planes
({\it correlated disorder}). Since it is obviously very important for technological
purposes to increase the critical current and optimize pinning, various methods
have been studied to increase disorder, such as electron irradiation,
which creates more uncorrelated point disorder, or heavy ion
irradiations, which creates parallel columns of defects (columnar
disorder).
Columnar defects have been found to be particularly efficient to pin the vortex lines
\cite{civale_columnar_prl,vanderbeek_columnar_long}
and their effect can be shown to be further optimized by deliberately
crossing them in splayed configurations \cite{hwa_splay_prl,hardy_splay_all}.
The physics of correlated disorder and its connections
to quantum disordered problems has also brought about many new and
interesting developments but goes beyond this paper (see e.g Refs
\cite{nelson_columnar_long,giamarchi_columnar_variat,blatter_vortex_review}).

Again, before the current interest in high
Tc materials, the dynamics of superconductors
was treated in an oversimplified way. Some of the main
experiments to be explained were (i) transport
experiments, i.e the shape of the $I$-$V$ characteristics
and (ii) magnetization relaxation experiments, i.e
the relaxation of the magnetic field profile inside a sample due to
thermally activated flux motion (flux creep). The motion far above
the threshold was supposed to be simple since the vortices slide very
fast, and average very well over the disorder.
One expects to find $v = F/\eta$, where
$\eta$ is the friction coefficient, $v$ is the average velocity
(and thus voltage) and $F$ the applied
force (proportional to the current $J$).
Motion around and below the threshold was described by
the conventional Anderson-Kim model, which
was sufficient to account for most of the observations
\cite{tinkham_book_superconductors}.
This was an effective
one particle model, where an unspecified piece of the vortex
lattice (a vortex bundle), moves as a single particle
in a one dimensional
potential energy landscape (tilted by the applied Lorentz
force). The potential could be chosen with some amount of disorder
or periodic because of the periodicity of the lattice. At $T=0$
this model immediately yields a critical force $F_c$ corresponding
to the maximal slope in the landscape. At $T>0$ thermal activation
allows motion by overcoming the
barrier $U_b$ needed to go from one minimum to the other. This
simple model yields:
\begin{equation}
v \sim \rho_{\rm ff} F e^{- U_b/T }
\end{equation}
At any finite temperature one would thus recover an exponentially small
but {\it finite} resistivity (i.e vortex mobility) $\rho \sim {\rm exp}[- U_b/T]$
(we recall that the current density is $J \sim F$,
the voltage drop $V \sim v$). This
mechanism, known as Thermally Assisted Flux Flow (TAFF) thus focuses
on the motion of individual vortices (or vortex bundles seen as a single
point). Thus in this conventional
approach the mixed phase is not a true superconductor at any finite temperature,
since it possesses a finite (albeit exponentially small) linear resistivity.
Also this model was insufficient to obtain any estimate for $J_c$,
which had to wait for more sophisticated approaches by
Larkin and Ovchinikov \cite{larkin_ovchinnikov_pinning}.
The simple TAFF approach became totally insufficient in high $T_c$
materials where giant thermal flux creep effects were soon observed.
It became apparent that at low temperature
the above TAFF law should be replaced by
\begin{equation}
v \sim \rho_{\rm ff} F e^{- U_b(J)/T }
\end{equation}
with an effective barrier $U_b(J)$ exhibiting a
strong dependence on the current, increasing rapidly with decreasing
$J$. Such a current dependent barrier could not be explained without a
more sophisticated theory taking into account the elasticity and
periodicity of the flux lattice.

A similar situation also occurred for the static properties of the
vortex lattice.  Paradoxically, although Larkin soon realized
that even weak disorder would have a strong impact, and lead to a
destruction of the
perfect translational order \cite{larkin_70}, such effects were not
investigated in detail until very recently.
In fact only after the discovery of high $Tc$
superconductors was it realized that disorder plays a crucial role.
Experimentally, it was observed very
early that the phase diagram is very different from the
predictions based on the ideal case (with only a melting
transition between a perfect solid and a flux liquid). Instead
there is an {\it irreversibility
line} \cite{malozemoff_irreversibility_line}
$H_{irr}(T)$ below which the system seems to exhibit a glassy
behavior, as can be seen from the history dependence of physical quantities, non
linear I-V with vanishingly small linear resistivity and
irreversibility. To understand the physical
properties of what was the solid phase, both from the point of view of
statics and dynamics, it is thus necessary to take disorder into
account.
Disorder will also affect the nature of the transition to the vortex
liquid phase.
The first experiments \cite{koch_vortex_continuous_transition}
concluded that the irreversibility line
corresponded to a {\it continuous transition}. However,
it became increasingly clear in later and more precise
experiments starting with observations
of jumps in the resistivity \cite{charalambous_melting_rc}, that
the transition at low fields and weak disorder
is in fact a {\it first order transition}.
The first order nature of the transition at low fields is by now well
established by a variety of techniques, such as transport
measurements\cite{charalambous_melting_rc,safar_tricritical_prl},
magnetization jump\cite{zeldov_diagphas_bisco} or specific heat
measurements \cite{schilling_heat_vortex}.

Thus in view of the new experimental results, the limitations and
inadequations of the conventional theories became clear. Both for
the technological applications of high-$T_c$ materials and from a purely
theoretical
point of view, it was thus of paramount importance
to understand the detailed properties of
the disordered vortex lattice.

\subsection{Different theoretical approaches}

Both experiments and analogies with other disordered systems in statistical
mechanics known to exhibit glassy effects, suggest
that the disordered vortex system too could lead to the
formation of a glassy state rather than that of a vortex lattice.
Although the nature of such a state was unclear,
one of its key properties should be, as in any glassy
system, to possess many low lying metastable states, with barriers between
them which diverge as a function of their ``separation'' in phase
space. Thus the low temperature phase, if it really is a ``true glass'',
should be characterized by the true vanishing of the linear resistivity
even at finite temperature \cite{fisher_vortexglass_short,feigelman_collective}.
Or if it is only a glass in an approximate way (with only finite barriers)
it should still lead to extremely small linear resistivity.
This is thus a significant departure from the
above models of thermally assisted
flux flow, which assumed {\it finite} barriers between pinned states.
A sign of an instability upon increasing disorder, presumably
towards a glass, was also found in the flux liquid
\cite{nelson_ledoussal_liquid}.

Divergences between
different theoretical approaches appeared in
the way of describing this glass phase.
Of course the full description is given by a disordered
Ginzburg-Landau functional. However such a theory is too complicated to
be analytically tractable. One way to proceed then is
to assume, as a phenomenological description,
that a complete destruction of the Abrikosov lattice
occurs even at very short length scales
\cite{fisher_vortexglass_short,fisher_vortexglass_long}.
Such an approach was prompted by an attempt to interpret the existing
decoration pictures at that time, and early experiments showing a
continuous phase transition.
It was then proposed \cite{fisher_vortexglass_short}, originally
by analogy with the Cardy Ostlund 2D disordered XY model
\cite{cardy_desordre_rg}
(a rather stretched analogy as it turns out - see Section~\ref{degal2}),
that there should be a low temperature phase, called the
``vortex glass'' with true zero linear resistivity.
To put some flesh on this rather
handwaving derivation of the physical properties,
it was later proposed \cite{fisher_vortexglass_long}
that the disordered Ginzburg-Landau model could be approximated
(while keeping the main ingredient of
Ref.~\cite{fisher_vortexglass_short}, namely the absence of a lattice)
by a simpler discrete XY model (in the superconducting
order parameter) with a quenched random gauge field.
As shown by numerical simulations
this ``gauge glass model'' leads to a zero linear resistivity
in $d=3$, a continuous ``vortex glass'' transition
(with scaling of the I-V curve near $T_g$). The physical
picture was thus of an effective ``spin glass'' order in the
superconducting order parameter.

The second approach is completely different. It retains
the elastic lattice structure at small scale \cite{feigelman_collective}
and describes vortex lines as strings having some elastic energy. Disorder is
then incorporated as acting directly on these elastic classical objects,
which amounts to forgetting about the phase of the order parameter
beyond the elastic theory. The vortex problem thus becomes a particular
case of the more general problem of an elastic system in the presence of
disorder.

Although different in nature, both theories agreed
that the disorder
essential to produce the glassy low temperature phase and the vanishing of
the linear resistivity, was also destroying at large scales the perfect flux lattice
existing in mean field theory.
The low temperature phase was therefore generally expected
to be a topologically disordered phase, lacking translational order.
Several calculations supported this point of view. Elastic
theory predicted at best a stretched exponential decay
of translational order
\cite{feigelman_collective,chudnovsky_pinning_long,bouchaud_variational_global}.
In addition general arguments tended to prove that
disorder would always favor the presence of dislocations
\cite{fisher_vortexglass_long,villain_cosine_realrg}.
The vortex lattice seemed to be buried for good.
However, although these approaches seemed to explain some aspects of the
problem, various others did not naturally fit in the framework
of these theories. As already mentioned, experimentally the transition
between the glass phase and the liquid is first order at low fields
\cite{charalambous_melting_rc,safar_tricritical_prl}
rather than continuous transition
observed at high fields which is predicted by the
gauge glass model. Furthermore, decoration experiments of
the flux lattice at very low fields (60 G) in several materials
were showing remarkably large regions
free of dislocations \cite{grier_decoration_manips} inconsistent with the
assumptions behind the ``vortex glass''.
Efforts to improve on the gauge glass model by incorporating screening
effects showed in numerical simulations \cite{bokil_young_vglass} that
the ``vortex glass''
phase would {\it not exist} in $d=3$ (the lower critical dimension
$d_{lc} >3$). There was also disagreement within a purely
elastic description: old calculations on the
related disordered elastic random field  XY model \cite{villain_cosine_realrg}
as well as more recent scaling arguments for the vortex lattice
\cite{nattermann_pinning} suggested, within a purely
elastic description, a slower, logarithmic growth of deformations.
All these problems, both theoretical and experimental, prompted for the
need of a quantitative theory of a disordered vortex lattice.
Before we look at it, let us further examine the
consequences of an elastic description of the vortex lattice.

\section{Statics of lattices with disorder}   \label{statsec}

Let us now follow the route of the elastic theory. Such an approach has
the advantage over the gauge glass approach to at least allow for some analytic
calculation, and is certainly a good starting point if the disorder is
weak. Of course the stability of the elastic approximation to
topological defects has to be (and will be) examined in the end.
Besides applying to vortex lattices,
such disordered elastic systems also cover many physical situations,
such as charge density waves \cite{gruner_revue_cdw},
Wigner crystals \cite{andrei_wigner_2d}
magnetic bubbles \cite{seshadri_bubbles_thermal,seshadri_bubbles_long},
Josephson junctions \cite{vinokur_josephson_short,balents_josephson_long},
the surface of crystals with quenched bulk or substrate disorder
\cite{toner_log_2} and domain walls in incommensurate solids
\cite{pokrovsky_talapov_prl}.
All these systems have in common a perfectly ordered
underlying structure modified by elastic distortions and
possibly by topological defects such as dislocations, due to
temperature or disorder. As for the vortex lattice, for many of such
systems the periodic structure
can be set in motion by an external force (e.g. and electric field for
the CDW or the Wigner crystal), and the velocity can be measured (e.g. by
measuring the current for CDW or Wigner crystal).

\subsection{Description in term of an elastic theory} \label{elasth}

Let us now look at the minimal model describing these different
physical systems. First, the pure
problem: we can ignore the internal structure of the objects (vortices,
magnetic bubbles etc.) and represent them as point-like objects. One can
then define an equilibrium position $R$. At that point one has to
distinguish between manifolds of internal dimension $d$
and periodic structures. For the manifold one
defines displacements $u$ (vectors of dimension $N$) relative to the
equilibrium position. A manifold is thus naturally embedded
in a space of dimension $D=N+d$ (for instance a $d=2$ interface
in a $D=3$ space). It has an elastic energy
\begin{equation} \label{real}
H_{{\rm el}} = \frac12 \sum_{\alpha,\beta}
\int \frac{d^dq}{(2\pi)^d}
u_\alpha(q) \Phi_{\alpha\beta}(q) u_\beta(-q)
\end{equation}
The $\Phi_{\alpha\beta}(q)$ is the elastic matrix. We can also
rewrite (\ref{real}) symbolically in real space as
\begin{equation}
H_{{\rm el}} \sim \frac12 \int d^dr~~ c~(\nabla u)^2
\end{equation}
where $c$ is the elastic coefficient (here isotropic for
notational simplicity).
For periodic systems the situation is more subtle. The equilibrium
positions are discrete $R_i$ (at least with respect of some of the
coordinates), and the lattice spacing is a genuine scale of the problem.
The elastic energy should depend on the discrete
differences between the displacements of two objects, such as
$u_i-u_{i+1}$. If the relative displacements
of two neighbors is small, a
situation realized at low temperature and small disorder,
one then usually takes the continuum limit by letting the
lattice spacing $a$ go to zero, or one performs a quadratic expansion
in $u$ to obtain an elastic energy similar to (\ref{real}), but
where the sum over $q$  is restricted to the first Brillouin zone.
In this continuous limit manifolds and periodic systems would thus
superficially look very similar from the point of view of their elastic
energy. However, this continuum limit should
be performed with great care since disorder can also vary at scales {\it
much smaller} than the lattice spacing $a$. Therefore the existence
of typical scale $a$ in the periodic structure must have some impact
on the physical properties of the system. As we will see it is
hidden in the expression of the density of objects as a function of the
displacements. Note that after averaging over disorder
only {\it relative displacements} have a direct
physical meaning since disorder is statistically translationally
invariant and one can obviously
translate the whole lattice without changing the elastic energy.

Two important quantities characterize the physics of such an elastic
system. The first one measures the relative displacements of two points
(e.g two vortices) separated by a distance $x$.
\begin{equation} \label{relative}
\tilde{B}(x)= \frac1N \overline{\langle [u(x)-u(0)]^2 \rangle}
\end{equation}
where $\langle \rangle$ denotes an average over thermal fluctuations and
$\overline{\qquad}$ is an average over disorder, and $N$ is
the number of components of $u$. The growth of $\tilde{B}(x)$ with
distance is a measure of how
fast the lattice is distorted. For thermal fluctuations alone in $d>2$,
$\tilde{B}(x)$ saturates at finite values, indicating that the lattice
is preserved. Intuitively it is obvious that in the presence of disorder
$\tilde{B}(x)$, will grow faster and can become unbounded.
$\tilde{B}(x)$ can directly be extracted from a direct imaging of the
lattice, such as performed in decoration experiments.

\begin{figure}
\centerline{\epsfig{file=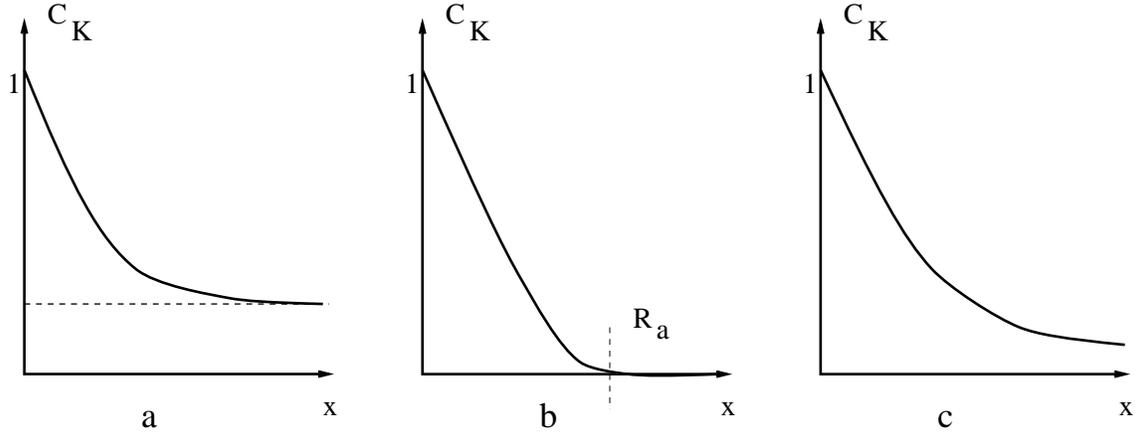,angle=-90,width=15cm}}
\caption{Various possible decays of $C_K(x)$. (a) For thermal
fluctuations alone $C_K(x)\to {\rm Cste}$, one keeps perfect $\delta$
function Bragg peaks, albeit with a reduced weight (the Debye-Waller
factor). (b) $C_K(x)$ decays exponentially fast. The structure factor
has no divergent peak any more, so translational order is destroyed
beyond length $R_a$, although some degree of order persists at short
distance. (c) $C_K(x)$ decays as a power law. The structure factor still
has divergent peaks but not sharp $\delta$ function ones. One retains
quasi-long range translational order. This is for example the case in
$d=2$ at small temperature (Kosterlitz-Thouless).
\label{fig1}}
\end{figure}

Related to $\tilde{B}(x)$, albeit different, is the structure factor of
the lattice, obtained by computing the Fourier transform of the density of
objects:
\begin{equation} \label{dens}
\rho(x) = \sum_i \delta(x - R_i -u_i)
\end{equation}
The square of the modulus $|\rho_k|^2$ of the Fourier transform of
(\ref{dens}) is measured directly in diffraction (Neutrons, X-rays)
experiments. For a perfect lattice the diffraction pattern consists of
$\delta$-function Bragg peaks at the reciprocal vectors of the lattice.
If some degree of short range order persists, individual peaks will still be present
although they might be broadened, and will not be simple $\delta$-functions
any more. The shape and width of any single peak is thus again
a measure of the degree of translational order in the lattice. To be
more quantitative, one can filter a single peak centered around the
reciprocal vector $K$ and Fourier transform it
back, to obtain a correlation function in real space. This correlation
function (called the translational order correlation function) is given by
\begin{equation}
C_K(x) = \overline{\langle e^{i K u(x)} e^{-i K u(0)} \rangle}
\end{equation}
Clearly the broader the peak the faster the decay of $C_K(x)$. $C_K(x)$
is therefore a direct measure of the degree of translational order that
remains in the system. Three possible cases are shown in
figure~\ref{fig1}.
For simple Gaussian fluctuations
\begin{equation} \label{approx}
C_K(x) = e^{-\frac{K^2}2 \tilde{B}(x)}
\end{equation}
but such a relation holds only qualitatively in general.

Of course one should also go beyond the simple elastic approximation
(\ref{real}) and worry about the possible existence of topological
defects such as dislocations. We will come back to this point later. If
unpaired dislocations are present they will destroy the translational
order exponentially fast beyond a certain length scale $R_D$ of the order
of the typical distance between such unpaired dislocations.

Although coupling to disorder depends on the precise microscopic aspects
of each system, it is usually possible to model it by pins put at
random positions in the system and coupling to the density.
\begin{equation} \label{complic}
H_{\rm rand} = \int d^dx V(x) \rho(x)
\end{equation}
The disorder potential is $V(x) = U_0 \sum_i \delta(x-x_i)$
where $x_i$ are the positions of the defects. The
$\delta$-function should be understood here as a
short range correlation function of range $r_f$. In
superconductors one has roughly $r_f \sim \xi_0$ which
can be much smaller than the lattice spacing $a$.
The effect of disorder depends
on the relative strength of the pins and the elastic forces. If the pins
are very strong, one must retain their discrete nature. On the other
hand if disorder is weak (for the regime of validity of such an
approximation see  \cite{giamarchi_columnar_variat}:
in $d \ge 2$ such an approximation always holds provided the strength of
each pin is weak), pinning occurs at
lengths much larger than the average defects distance (this notion
will be made more quantitative in section~\ref{simplesec}).
It is then legitimate to replace $V$ by a simple Gaussian potential with
a correlator
\begin{equation} \label{cordisorder}
\overline{V(x)V(x')} = \Delta(x-x')
\end{equation}
where $\Delta$ is a function of range $r_f$.
(\ref{real},\ref{complic}) describes the most general elastic system
coupled to disorder.
Even with the Gaussian disorder $V$ (\ref{real},\ref{complic})
is a rather formidable theory to solve because of the highly non-linear
nature of the coupling to disorder.

Before attacking the problem with the full force of replica theory and
 renormalization group,
let us examine some simple arguments to understand its physics.
This is instructive, since a good part of the physics can be derived
simply, but also shows how too simple pictures can also
prove misleading when used beyond their range of validity.

\subsection{Larkin model} \label{simplesec}

Since even Gaussian disorder is too complicated to be studied directly,
Larkin had the remarkable idea \cite{larkin_70} to replace
the coupling to the random potential (\ref{complic}) by random forces
acting on the vortices and coupling directly to the displacements
\begin{equation} \label{larkin}
H_{\rm L} = \int dx f(x) u(x)
\end{equation}
where $f$ has Gaussian correlations. Being linear
(\ref{real}-\ref{larkin}) is now rather straightforward to solve. It
yields a displacement correlation function of the form
\begin{equation} \label{larkinb}
\tilde{B}(x) = l^2 (x/R_l)^{4-d}
\end{equation}
where $l$ is any lengthscale and $R_l$ the distance for which the
relative displacements are of order $l$.
Since the theory is simply Gaussian, $\tilde{B}$ and $C_K$ are
simply related by (\ref{approx}). Disorder thus destroys the long range
translational order below four dimensions, and leads to an exponential
decay of the correlation function in the physically relevant $d=3$
case. It is important to note that the Larkin model has {\em two}
characteristic length scales:
\begin{itemize}
\item
$R_c$ which is the distance in the XY-plane at which the relative
displacements of the flux lattice are of order the {\em correlation length}
of the random potential, $r_f$ (which is of order $\xi_0$ at low
temperature), and
\item
$R_a$, which is larger, and is the distance over which the relative
displacements of the flux lattice are of order the {\em lattice spacing}
$a$.
\end{itemize}
$R_a$ is related to the disorder strength by (\ref{radef}) below, and
$R_c$ is given by a similar expression but with $a$ replaced by $\xi_0$.
Using (\ref{approx}), $R_a$ can be related to the length beyond which
translational order is really destroyed. The naive physical picture emerging
from this model is the one of a crystal breaking into crystallites of
typical size $R_a$, as schematized in case (b) of figure~\ref{3regimes},
and would correspond for
the translational correlation function to case (b) of figure~\ref{fig1}.

However this model has several limitations built in the
approximation (\ref{larkin}) of the random term. Since it
lacks the translation symmetry $u\to u+a$ of the physical model
(\ref{complic}), it gives a real meaning to the displacement field $u$
itself, whereas the original model can of course be sensitive only to
the {\it position} of the line itself. The Larkin model can thus be
viewed as an expansion in powers of the
displacements $u$, and only holds when the displacements are small.
That the approximation (\ref{larkin}) is too crude is also apparent when
one tries to determine the pinning force.  Since the Larkin Hamiltonian
is {\it linear} in $u$, any global translation of $u$ does not change
the average energy, and thus the pinning force is zero. In order to
describe correctly the pinning one {\it needs} the nonlinearities of
(\ref{complic}).

In a masterful stroke of physical intuition, Larkin and Ovchinikov (LO)
realized \cite{larkin_ovchinnikov_pinning} that the breakdown of
validity of the Larkin model occurred
exactly at the lengthscale corresponding to the critical pinning force.
Indeed one can associate to an applied force $F$, a lengthscale in the
static problem corresponding to the size of the smallest bundle moving.
This size can be obtained by balancing the elastic plus pinning energy
with the Lorentz force work. Since the pinning energy grows only as
$L^{d/2}$ and not as the volume, a small external force will be able to
move only large bundles (we will come back to this point in
section~\ref{charlength}), thus the
smaller the force, the larger the size of the bundle. As long as
the Larkin model is valid, no pinning exists and the lattice flows
``freely'' and one should be {\it above} $F_c$. It is only below $F_c$,
i.e. for length scales larger than those for which the Larkin model
applies that one can expect anomalous transport. One may think naively
that the expansion breaks down at $R_a$. In fact this occurs much
sooner at $R_c$, as was
realized by LO: the expansion in displacements becomes incorrect as soon
as a line can be considered off its equilibrium position, i.e. when it
has moved by more than its intrinsic width roughly given by the core
size $\xi_0$. More precisely the expansion breaks down at scales for
which the allowed relative motion is such
that the random potential cannot be approximated by its
slope only. The critical force is thus associated to the typical
energy for which motion of the lines is of order $r_f \sim \xi_0$.
If $R_c \gg a$, this energy is obtained by balancing the energy gained
due to the
applied force $\sim u R_c^d F_c$ to the typical pinning energy
$E_{\rm pin}$ of
a bundle of size $R_c$. Such a bundle has an elastic energy
$E_{\rm el}\sim c R_c^{d-2} u^2$ and a pinning energy
$E_{\rm pin} \sim R_c^{d/2}\Delta^{1/2} u$ of the same order
$E_{\rm el}\sim E_{\rm pin}$ for
displacements $u\sim r_f$. The critical force density is thus given by
\begin{equation} \label{pinningf}
F_c = \frac{c r_f}{R_c^2}
\end{equation}
Formula (\ref{pinningf}) is quite remarkable since
it makes it possible to obtain one of the most interesting physical
quantities for the dynamics
directly out of length scales of the {\it static} problem. It also
shows that that simple TAFF arguments are too simple: in
order to correctly describe the effect of the external force and hence
the motion, the {\it collective} behavior of the lines should be
investigated. Indeed the motion of individual lines would lead to
a cost in energy much too high compared to the energy
gained due to the motion.
On the other hand, the motion of large enough bundles are able to
overcome the pinning force. Determination of the transport below $F_c$
thus requires the knowledge of the static properties of the lattice at
lengths larger than $R_c$, for which a solution of the nonlinear
problem (\ref{complic}) is required.

\subsection{Characteristic lengthscales}  \label{charlength}

In order to gain a physical insight in the static properties of
the disordered lattice one can use simple dimensional analysis.
In the presence of many weak pins, $u$ cannot
adapt to take advantage of each of them, due to the cost in elastic
energy. One can assume that $u$
varies of $\sim a$ over a length $R_a \gg a$. The density of kinetic
energy is $\sim c(a/R_a)^2$, where $c$ is an elastic constant.
For the disorder (\ref{complic}), one has to be very careful to separate
its various Fourier components. Indeed the period of lattice introduces
a {\it natural} scale $a$. The Fourier components of the disorder close
to vector
of the reciprocal lattice thus play a special role, as appears when one
rewrites the density (such a decomposition is {\it exact} in the elastic
limit\cite{giamarchi_vortex_short,giamarchi_vortex_long})
\begin{equation} \label{transparent}
\rho(x) \simeq \rho_0 (1 - \partial_\alpha u_\alpha (x) +
        \sum_{K \ne 0} e^{i K x} e^{-i K \cdot u(x)})
\end{equation}
The gradient term is simply the standard change of density due to a
compression of dilatation, whereas the other terms correspond to a
translation of the lattice without any change in the average density.
This decomposition of the density is very similar to the density
modulation in a CDW, where the role of $u$ is played by the phase of the
CDW, and only one $K$ vector exists.

If one assumes that $u$ is roughly constant over a length $R_a$, the
Fourier components of the disorder with components close to $e^{i
K x}$ will give different contributions.  It is easy to see that the
long wavelength part of the disorder can only be relevant for $d \leq 2$, and thus can
be safely dropped. The main contribution comes from the Fourier modes of the disorder
which have a
periodicity close to the one of the lattice, and which do not tend to change the local
density but rather to shift the lattice locally:
\begin{equation} \label{disorder}
H^{{\rm dis}}_{q \sim K}
 = \rho_0 \int d^dx V(x) e^{i K x} e^{-i K u(x)}
\end{equation}
This sum can be viewed as a random walk in the complex plane
\cite{fukuyama_pinning} and
the value of $u$ adjusts itself to match the
phase of the random potential. Therefore the gain in
energy density due to the disorder term is of order
\begin{equation}
H^{{\rm dis}}_{q \sim K} \sim \Delta_K^{1/2} /R_a^{d/2}
\end{equation}
where $\Delta_K$ are the Fourier modes
of the disorder correlator (\ref{cordisorder}), for vectors of the
reciprocal lattice.
Minimization gives  a length $R_a$ of the order of
\begin{equation}  \label{radef}
R_a  \sim  a\left(c^2 a^{d}/\Delta_K\right)^{1/(4-d)}
\end{equation}
Weak disorder is thus relevant below $d=4$. Beyond the length $R_a$ the
relative displacements are larger than the lattice spacing $a$. One can thus
naively see the lattice as broken into domains of size $R_a$ above which
the translational order would be destroyed (displacements are larger
than $a$). This image would be in agreement with the one one gets from
the Larkin model (used {\it beyond} its range of validity), and as such
has been often accepted as correct. As we will see this vision is too
naive, but the length $R_a$
is indeed one important lengthscale of the problem. Way below $R_a$, one
could also identify $R_c$ with the length for which the displacements
are of order $\xi_0$. The same analysis, but with $u\sim \xi_0$ would
give $R_c  \sim  \xi_0\left(c^2 \xi_0^{d}/\Delta_K\right)^{1/(4-d)}$.
It is important to note that the above arguments give only the way these two
lengths depend on disorder. Obviously $R_c < R_a$ and in fact
one can see that $R_c/R_a \sim (a/\xi_0)^{1/\nu}$, where $\nu$ is the
roughening exponent defined by $B(x)\sim x^{2\nu}$, or equivalently by
saying that the displacement $u$ scale as $u\sim L^\nu$. For the
Larkin model (\ref{larkinb}) implies $2\nu=4-d$.
Two kinds of domains
can thus be defined. For length smaller than $R_c$ the Larkin model is
valid and the displacements grow as $x^{(4-d)/2}$ giving
(\ref{larkinb}). Although this behavior can be extracted from an exact
solution of the Larkin Hamiltonian it can also be obtained
by simple dimensional analysis. If one assumes that $u\sim
L^\nu$, balancing the elastic and disorder terms in the Larkin
Hamiltonian gives
\begin{equation}
L^{d-2 + 2 \nu} \sim L^{d/2 + \nu}
\end{equation}
thus yielding $u \sim L^{(4-d)/2}$.
Note that $R_c$ is in fact the {\it only} length that can be extracted
from the Larkin model and that gives the bundle size corresponding to
$F_c$.
It is important to carefully distinguish between $R_c$ and $R_a$, since
they correspond to two physically different lengthscales. Serious
confusion exists in the literature in this respect (see e.g.
\cite{yaron_neutrons_vortex,giamarchi_vortex_comment}).

What happens beyond the LO length $R_c$ is more subtle and has only been
understood recently. Since the lines have moved by more than their
intrinsic width, one cannot use Larkin's random force approximation, and
the full nonlinear Hamiltonian (\ref{complic}) should be used.
Intuitively one can understand this regime provided that the
displacements remain small compared to the lattice spacing, i.e. for
separations smaller than $R_a$. Indeed in that case each line sees it
{\it own} realization of the random potential. One is thus led back to
the random manifold problem, where each point of the manifold sees an
independent random potential, since it corresponds to different internal
manifold coordinates. To obtain a simple estimate for the growth of
displacements one can again balance the elastic and random energy, but
this time keeping the potential term (\ref{complic}). Similar scaling
arguments give a mean-field (Flory) roughening exponent $u \sim
L^{\nu_F}$ with $\nu_F=(4-d)/(4+N)$ ($d$ being the space dimension and
$N$ the number of components of the displacement field as defined in
section~\ref{elasth}). This exponent is
not exact but a good approximation of the true random manifold exponent
$\nu_{\rm rm}$.
Contrary to what happens in the Larkin regime, the random manifold
regime corresponds to a glassy regime, where the system can find many
metastable states.

A manifestation of this glassy regime can be found in
transport, since as already mentioned the detailed transport properties
depend on how the lattice is pinned. To estimate the $v-F$
characteristics one can use the near equilibrium collective creep
arguments \cite{feigelman_collective}, balancing the
energy gained due to the external force with the energy of the pinned
system. If the system is in the ground state the closest metastable
state of a bundle of size $L$ scales as $L^\theta$
(where $\theta$ is the energy exponent $\theta=d-2+2\nu$).
The standard assumption is that energy barriers scale with the
same exponent $\psi=\theta$ (it was possible in some cases to check it
explicitly). In order to
reach this metastable state the energy loss should equal the gain due to
the external force giving
\begin{equation}
L^\theta \sim L^d L^\nu F
\end{equation}
The smallest bundle to move will have a size $(1/F)^{1/(d+\nu-\theta)}$
and an energy $E = (1/F)^{\theta/(d+\nu-\theta)}$. The velocity will be
non-linear and given by
\begin{equation} \label{creeplaw}
v \sim e^{-\frac{U_c}{T} (F_c/F)^{\theta/(d+\nu-\theta)}}
\end{equation}
where $U_c$ is the barrier at scale $R_c$.
The existence of many metastable states separated by diverging barriers
thus manifests itself in the vanishing of the {\it linear} resistivity.
This is different from the naive TAFF picture (which corresponds to
finite barriers). Thus one of the main achievements of the collective
creep picture is to account for the existence of true superconductivity.
For that the needed ingredients were elasticity and disorder.
It is interesting to note that the Anderson-Kim model can be
generalized \cite{ledoussal_anomalous_diffusion} (and solved
exactly) to an arbitrary one dimensional landscape. One can then
recover a lot of the features of these glass phases, and
of the vortex glass transition by choosing a landscape
with {\it long range correlations} which captures the diverging barriers
in a phenomenological way.

However even below $R_a$ the validity of this elastic approach was
questioned. Indeed the entire above study completely ignores the
possibility of creating topological defects, either in the statics or
for the dynamics. These defects could ruin the above nice result
(\ref{creeplaw}) and the very existence of a true glass phase.
In the absence of disorder and at low temperature it
is of course obvious that defects able to destroy the translational
order, such as unbounded dislocations, cannot appear. In the presence of
disorder, the issue was less clear. A long line of arguments, going
back to \cite{villain_cosine_realrg} and further developed in
\cite{fisher_vortexglass_long} were put forward to indicate that
in the presence of disorder, however weak, dislocations
should always proliferate. Since the elastic theory already
``showed'' that long range order should be destroyed beyond $R_a$
these arguments were not challenged in subsequent studies which
addressed the question of the behavior of the
elastic deformations \cite{nattermann_pinning,blatter_vortex_review}

As we will show such arguments were incorrect, though worth examining.
An Imry Ma type argument is the following.
The core energy cost of a dislocation cannot be avoided and scales as
$L^{d-2}$.
A dislocation loop of size $L$ creates extra-displacements of order
O(1) up to
logarithms,
in a region of size $L^{d}$. By adjusting the position of the loop
one can hope to gain an energy from disorder $L^{d/2}$. Thus below $d=4$
large (infinite) dislocation loops will be favorable.
The argument is flawed because it is again implicitly assuming
the same physics as in Larkin's random force
model for which the disorder energy is linear in the displacement. For
the real model (\ref{disorder}) the energy varies as $\cos(K u)$;
adding a dislocation displacement will not necessarily gain enough disorder
energy. We will come back to this point and give the correct arguments
in section~\ref{dissec}. The issue of the existence of dislocations is
an important one. Indeed
if dislocations are generated by disorder, the elastic theory does not
hold and its conclusions are questionable. In particular, if
dislocations are present the non-linear $v-F$ characteristic should be
replaced by a TAFF characteristic due to the plastic deformations.

\subsection{Full solution of the periodic problem} \label{dissec}

Let us now examine the full solution of the problem, and in
particular examine the physical properties beyond the
length $R_a$ for which the {\it periodicity} of the lattice becomes
important. To do so we use a variational formulation of the
problem
\cite{giamarchi_vortex_short,giamarchi_vortex_long,%
korshunov_variational_short}.
Similar results have been obtained using the functional renormalization
group approach \cite{giamarchi_vortex_short,giamarchi_vortex_long}.

\subsubsection{A variational formulation}

In order to average over the disorder in (\ref{real},\ref{complic}), we
use the replica trick and obtain
\begin{equation} \label{cardyos}
H_{{\rm eff}}  =  \frac{c}2 \int d^dx (\nabla u(x))^2
 - \int d^dx  \sum_{a,b} \sum_{K\ne 0} \frac{\rho_0^2\Delta_K}{2T}
\cos(K \cdot (u^{a}(x)-u^{b}(x)))]
\end{equation}
of course the full elastic Hamiltonian should (and has) been used, the
above being a simplified notation. In particular for vortex
lattices the anisotopy introduced by the magnetic field between the
in-plane and along the field directions can be trivially treated by a
rescaling (for more details see \cite{giamarchi_vortex_long}).
We now look for the best
trial Gaussian Hamiltonian
$H_0$ in replica space which approximates (\ref{cardyos}). It has the
general form \cite{mezard_variational_replica}
\begin{equation} \label{variat}
H_0 = {1 \over 2} \int {d^dq \over (2 \pi)^d} G^{-1}_{ab}(q)
u_a(q) \cdot u_b(-q)
\end{equation}
where the $[G^{-1}]_{ab}(q)$ is a $n$ by $n$ matrix of variational parameters.
We obtain by minimization
of the variational free energy $F_{{\rm var}}=F_0+\langle H_{eff}-H_0
\rangle_{H_0}$ the saddle point equations for the variables $G^{-1}$.
The technical details of the solution can be found in
\cite{giamarchi_vortex_short,giamarchi_vortex_long}. Such an
approximation is expected to be a good one unless kink excitations
around the pinned configurations are very important. As can be
confirmed by an independent functional renormalization group calculation
\cite{giamarchi_vortex_short,giamarchi_vortex_long}
(in $d=4-\epsilon$ dimensions), the variational approach seems to capture here
the correct physics.

Two general classes of solutions can exist for (\ref{variat}). One
preserves the symmetry of permutations of replica, and amounts to mimic
the distribution (thermal and over disorder) of each displacement mode $u(q)$
by a simple Gaussian. The other class, which is a better approximation in the glass phase,
breaks replica symmetry and approximates effectively the distribution of displacements
by a hierarchical superposition of Gaussians centered at different
randomly
located points in space according to a Parisi scheme
which is described in detail in \cite{mezard_variational_replica}.
As can be expected the most stable solution is the one that
breaks replica symmetry (a full RSB for $2<d<4$), confirming the
glassy properties.  On the variational solution the two lengthscales
$R_c$ and $R_a$ also appear and define three regimes as a function of
the separation $r$ which will be discussed below. In the present
problem the physics contained in the RSB solution can be expressed
as follows: each Gaussian at the lowest level of the
hierarchy is associated to a different metastable ``pinned'' position of
the manifold, corresponding to the Larkin length $R_c$.
Let us illustrate it, for simplicity, on the simplest case of a one step RSB
solution, which is the solution in $d=2$ (we thus anticipate
on section \ref{degal2}). The double distribution over environment
and thermal fluctuations is approximated as follows. In each environment
there are effective ``pinning centers'' corresponding to the
low lying metastable states (prefered configurations). Since all $q$ modes
are in effect decoupled within this approximation, for each $q$ mode
a prefered configuration (a state) is $u_\alpha(q)$. They are
distributed according to:
\begin{eqnarray}
P(u_\alpha(q)) \sim \prod_{\alpha} e^{ -\frac{c}{2 T_g} q^2 |u_\alpha(q)|^2 }
\end{eqnarray}
Each is endowed with a free energy $f_\alpha$ distributed
according to an exponential distribution $P(f) \sim \exp(u_c f/T)$
(here $u_c=T/T_g$).
Once these seed states are constructed,
the full thermal distribution
of the $q$ mode $u_q$ is obtained by letting it fluctuate thermally
around one of the states:
\begin{equation}
P(u_q) \sim \sum_{\alpha} W_\alpha
e^{-\frac{c}{2 T} (q^2 + R_c^{-2}) |u_q - u_\alpha(q)|^2}
\end{equation}
where each state is weighted with probability
$W_\alpha=e^{-f_\alpha/T}/\sum_\beta e^{-f_\beta/T}$.

One thus recovers qualitatively the picture of Larkin Ovchinikov as the solution
of the problem with the replica variational method. The LO length
naturally appears as setting the (internal) size of the elastically
correlated domains. The full RSB case corresponds to more level in this hierarchy of
Larkin domains (in some sense there are clusters of domains of
size larger than $R_c$) and the way this hierarchy scales with
distance reproduces the Flory exponents for displacements and
energy fluctuations.

The other important method which provides a picture consistent
with this one is the functional renormalization group FRG
developed by D.S. Fisher \cite{fisher_functional_rg,balents_frg_largen}.
There it is found that beyond $R_c$ a non-analyticity develops in the
coarse grained renormalized disorder correlator. This corresponds
to the renormalized random potential developing cusp singularities
\cite{narayan_fisher_cdw}
(consistent with the LO picture of the medium breaking into
Larkin domains). The FRG has the advantage to take better non linearities into account,
but it does work only near $d=4$. The variational method on the other hand
works in any dimension as an approximation. Since it is a
Hartree replica method it does becomes exact
when $N \to \infty$ for the manifold problem. In the large $N$ solution
the various pure states $u_\alpha$ do not talk to each other (there is
true breaking of ergodicity). Presumably this should be improved
by instanton type contributions for realistic finite $N$, though
how to do this remains a totaly open question.
Even if taken simply as an approximation
and with a grain of salt
the GVM gives however a very reasonable physical picture.

\subsubsection{Full solution: the three regimes}

For point-like disorder, there are
{\it three} different regimes.
The variational approach predicts the full crossover function between
three regimes. There are as follows:

i) When $\tilde{B}(r)$ is shorter than the square of
the Lindemann length $l_T^2 = \langle u^2 \rangle$, the
thermal wandering of the
lines averages enough over the random potential and the model becomes
equivalent to Larkin's model for which $\tilde{B}(r) \sim |r|^{4-d}$. At
low temperature, $l_T$ is replaced by the superconducting coherence
length $\xi_0$ (i.e. the
correlation length of the random potential \cite{feigelman_collective,%
bouchaud_variational_vortex}).
At zero temperature it equals the length defined by
$R_c$ defined by the Larkin-Ovchinikov length
and in general $l$ can be thought of as the Larkin Ovchinikov length renormalized
by temperature.
Below this length the elastic manifold
sees a smooth potential with well defined derivatives, thus
a local random force can be defined.
Indeed expanding in $u$ the disorder
potential energy in (\ref{disorder}) gives a random force term $f.u$
with
$f(x)= \sum_K V(x) K exp(-i K x)= \nabla V(R_i)$. In the sum over harmonics
the maximum $K$ is $K_{max}=2 \pi/\xi_0$. Thus
this expansion is valid only as long as $u \ll \xi_0$. This defines the
range of validity of the Larkin regime, i.e at $T=0$ $x < R_c$
and more generally $x<l$. Of course this first regime only exists if
$R_c > a$. From the point of view
of the variational solution, this regime corresponds to a replica
symmetric part, consistent with the fact that there are no metastable
states (and thus no pinning in Larkin's random force model).

ii) For $l_T^2 \ll \tilde{B}(r) \leq a^2$,
$\tilde{B}(r) \sim r^{2\nu}$ where $\nu\sim 1/6$:
this is the random manifold regime mentioned above where
each line sees effectively an independent random potential.
This can be seen, on a more mathematical level,
from our model by summing over all the harmonics
for instance on the replicated Hamiltonian (\ref{cardyos}). One gets
$V(u) \sim \sum_{R_i} \delta(u_a - u_b - {R_i})$. For $u \ll a$
only the $R=0$
term contributes and each line sees an independent random potential.
This intermediate random manifold regime holds up to the length $R_a$
such that $\tilde{B}(R_a) \approx a^2$ at which periodicity becomes
important. It is noteworthy that for models for which only one harmonic
exists, such as for CDW, the random manifold regime does not exist,
and one directly crossovers from the Larkin regime to the asymptotic
regime iii).
In the RM regime replica symmetry is fully broken, a signature of the
various metastable states and of pinning.

iii) For $r>R_a$,
the periodicity of the lattice becomes important and
$\tilde{B}(r) \sim A_d \log|r|$ where $A_d$ is a universal
amplitude depending on dimension only. To check the result of the
variational method we also computed $A_d$ using a functional
renormalization group procedure \cite{fisher_functional_rg},
in a $\epsilon = 4-d$ expansion. These two rather
different methods agree at order $\epsilon$ within 10\%.
Within the Gaussian approach, (\ref{approx}) gives
the translational correlation function $C(r)$ which
has a slow algebraic decay in $d>2$, $C(r) \sim (1/r)^{ A_d}$
and quasi-long range order persists.  This is
a reasonable lower bound for $C_K(r)$. It may give
the exact asymptotic decay or it is also possible that atypical ``return to
the origin'' events (i.e a singularity at $u=0$ of the scaled probability of $u$)
could make this decay {\it slower}.
A similar situation is discussed in \cite{mitra_pld_rmn}.

\begin{figure}
\centerline{\epsfig{file=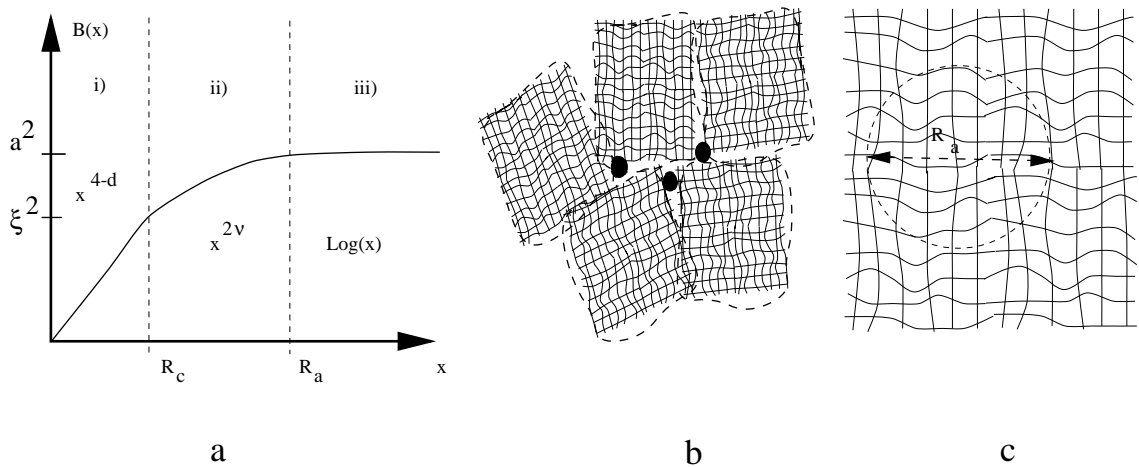,angle=-90,width=15cm}}
\caption{a) The generic behavior of the relative displacements
correlation function $B(x)$ for any disordered elastic system, with the
three regimes described in the text. As explained in the text depending
on the values of $R_c$ or $R_a$ regime i) or ii) might be absent. At
large distance displacement grows only logarithmically and quasi-long
range translational order exists.
b) shows the {\it incorrect} but commonly believed view
of a disordered elastic system. $R_a$ would correspond
to a length above which translational order is destroyed and the system
would break into ``cristallites''. Topological defects (represented as
black dots) would be generated due to disorder. c) is the {\it real}
situation for weak disorder (the Bragg Glass):
$R_a$ is just the crossover scale for which relative
displacements are of order $a$, and above which they grow very slowly.
So although displacements can become large, the system preserves
quasi-long range translational order, and no topological defect exists.
\label{3regimes}}
\end{figure}

The above regimes describe {\it generically} a disordered periodic elastic
system. A summary is shown in figure~\ref{3regimes} with the main
characteristic scales.
Quite unexpectedly {\it periodicity} helps the system to keep
quasi-long
range translational order ! $R_a$ is not a sort of crystallite size
beyond which translational order would be lost, but on the contrary
the length above which displacements grow {\it more slowly}.
The difference between these two physical pictures is illustrated in
figure~\ref{3regimes}.
This near saturation of the displacements can be understood by noting
that the system wants to minimize its total energy due to disorder. A
single line thus cares little in making displacements much larger
than $a$ since it would merely try to ``steal'' the random potential of
one of its neighbors, with little gain for the total energy and at a
huge elastic cost. Periodicity has thus drastic consequences and
paradoxically leads to a more ordered situation than for a simple
manifold.

Using the variational methods, many other results can be derived ranging
from the full crossover function $\tilde{B}(x)$ to the detailed behavior of the
translational correlation function $C_K(r)$ and we
refer the reader to
\cite{giamarchi_vortex_short,giamarchi_vortex_long} for further
details.

\subsubsection{Dislocations or no dislocations}

This striking result that quasi-long range order survives
has been derived within an elastic theory, assuming the absence of
dislocations. The alleged importance of dislocations in a disordered system
\cite{fisher_vortexglass_long} makes it mandatory to further investigate
carefully whether dislocations can modify the above
result. In fact, using an energy argument
\cite{giamarchi_vortex_long}, one can get convinced that dislocations
are much less relevant than commonly assumed.
The argument is as follows: the core energy of a dislocation loop
of length $L$ grows
as $L^{d-2}$. Since a dislocation involves $1/r$ displacements around
its core, the
total cost associated with an unbound dislocation is therefore $L^{d-2}
\ln(L)$. Now the cost of an elastic deformation which can be typically
relaxed by allowing for a dislocation loop is, provided $u\sim
L^\nu$ of the order of $L^{d-2 + 2\nu}$. So if translational order is
destroyed ($\nu > 0$), i.e. if the Larkin or the random manifold regime
were true up to infinite scales, it would indeed be favorable to create
dislocations. However if quasi long range order persists $\nu=0$ and for
weak disorder, the cost of a dislocation would always be higher for weak
disorder than the one of an elastic deformation, and dislocations are
{\it not} generated by disorder. The elastic solution is thus {\it self
consistently stable}. This implies self-consistently the existence
of a {\it thermodynamic glass phase}, as far as energy and very low
current transport properties are
concerned, retaining a nearly perfect (i.e. algebraic) translational order.
Since this phase exhibits Bragg
peaks very much like a perfect lattice we christened it the ``Bragg
glass''.

The prediction \cite{giamarchi_vortex_long} that a phase
{\it without} topological defects should be stable
at weak disorder, which also applies to the random field
XY model, received subsequent further support both from numerical
simulations \cite{gingras_dislocations_numerics,ryu_diagphas_numerics},
analytical calculations in a layered geometry
\cite{kierfeld_bglass_layered,carpentier_bglass_layered}
and, very recently, a proof using improved scaling and energy arguments
\cite{fisher_bragg_proof}.

\subsubsection{Bragg glass and other glasses}

Because of their original periodicity, periodic systems in a random
potential have a radically different physics than originally
expected: quasi-long range order and no topological defects !
Of course, since it retains a ``lattice'' structure
and Bragg peaks, this glass phase is widely differs from the vortex
glass picture based on a random gauge model. It is also different from the
naive original picture based simply on elastic manifolds. Indeed,
as one sees on on figure~\ref{3regimes} if the random manifold regime
had survived at large scales (beyond $R_a$) the same energy argument
implies that dislocations {\it would have been generated}
spontaneously (since there
$\nu >0$). The theory of the Bragg glass is thus poles appart
from the previously proposed theories for the glass phases of
superconductors.
Although we have insisted here on the vortex aspects of this phase
relevant mostly
to vortex systems let us
emphasize again that it is quite generic to any elastic disordered
system and should be observable in other situations. In
particular, CDW systems for which direct measurements of $C_K$ is
possible by x-ray diffraction should be good candidates.

For the vortex problem the Bragg glass has of course many
experimentally observable consequences. In particular,
since such a phase is nearly as good as a perfect lattice as far
as translational order is concerned, it is natural to expect it to melt
through a first order phase transition. We proposed
\cite{giamarchi_vortex_long} that the phase seen experimentally
at low fields in type II superconductors
was in fact the Bragg glass, solving the
apparent impossibility of a pinned solid.
This allowed to account naturally for the first-order transition
and the decoration experiments.
Neutron experiments (measuring directly $C_K$) can be naturally
interpreted in term of the Bragg glass
\cite{yaron_neutrons_vortex,giamarchi_vortex_comment}.
For a detailed discussion of the
various experimental consequences we refer the reader to
\cite{giamarchi_vortex_long,giamarchi_diagphas_prb}.

\begin{figure}
\label{figure1}
\centerline{\epsfig{file=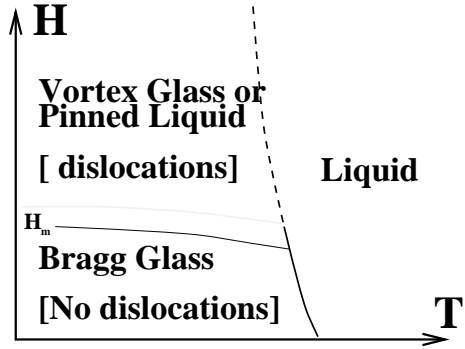,angle=-90,width=6cm}}
\caption{The stability region of the Bragg glass phase in the magnetic
field $H$, temperature $T$ plane is shown schematically. The thick line
is expected to be first order, whereas the dotted line should be either
second order or a crossover. Upon increasing
disorder the field induced melting occurs for lower fields as
indicated by the thin solid line. \label{phasediag}}
\end{figure}

But one of the most interesting consequences is provided by the
constraints that the mere {\it existence} of a Bragg glass phase
imposes on the $H-T$ phase diagram of superconductors. Indeed since the
Bragg glass should {\it not} contain unbound topological defects, a
phase transition should
exist towards {\it another} phase containing topological defects when
disorder is increased (an upper bound for the limit of stability for the Bragg glass
is of course $R_a \sim a$). One can show that in the range of fields
relevant for most high $T_c$ superconductors, increasing the field is
equivalent to increasing the disorder. Thus the simple existence of the Bragg
glass imposes \cite{giamarchi_vortex_long} that a transition in field
should exist. The nature of the high field phase is still unclear
both experimentally and theoretically. It
could be either a pinned liquid or another glass
(the putative ``vortex glass'' ?). What is clear is that
due to the presence of topological defects in that other phase
one expects it to melt in
a much more continuous fashion into the liquid (or simply to undergo a
crossover), and thus may be consistent with the observed
continuous (or gradual) transition at high fields.
These considerations led us to propose the phase diagram depicted
schematically in figure~\ref{phasediag}.
Let us emphasize again that the existence of such a transition in field
is the direct consequence of the fact that the low field {\it
thermodynamic} phase has no topological defects.

Several recent experiments can be interpreted to confirm the picture
proposed in \cite{giamarchi_vortex_short,giamarchi_vortex_long}.
In BSCCO neutron diffraction peaks are observed at low fields and
disappear upon
raising the field \cite{cubbit_neutrons_bscco}. The phase diagram of BSCCO
\cite{zeldov_diagphas_bisco} is also compatible with our theory, the
second magnetization peak line corresponding to the predicted
field driven transition. This line is found to be relatively
temperature independent
at lower temperatures and to be shifted downwards upon increase of
point disorder
\cite{khaykovich_diagphas_bisco,chikumoto_secondpeak_bisco}.
Similar types of phase diagrams are also observed in a variety of
materials, including YBCO, organic superconductors and heavy fermion
compounds, which seem to indicate that this is indeed a quite generic
mechanism. More experimental consequences and references can be
found in \cite{giamarchi_diagphas_prb}. More experimental work
will be needed to confirm the proposed picture.

On a theoretical level, going beyond the simple topology of the
phase diagram proves for the
moment to be very difficult, since we are impaired by our lack of
knowledge of the high field phase. Describing such a phase
would mean to be able to
treat a disordered elastic system in which there is also a finite amount
of topological defect, a quite challenging problem, but at present
beyond our reach. We even lack tools to obtain semi-quantitative
estimates for the positions of the melting lines. A blind application of
a Lindemann criterion to the transition in a field gives
a numerical estimate \cite{giamarchi_diagphas_prb} of the melting field
$H_M$ for BiSCCO of
$H_M\sim 400 G$ in good agreement with the observed experimental values
\cite{khaykovich_diagphas_bisco}. However it is far from clear that the
Lindemann criterion which works indeed quite well to describe thermal
melting, can also capture the physics of this peculiar disorder-induced
melting transition. It could however give correctly
the value of the field at the transition as can be checked for special
geometries where the field transition can be computed. Clearly, dealing
with a disordered elastic system in presence of a {\it finite} density
of topological defects (the only way to really describe either the
melting transition or the transition in field), is the next challenge !

\subsection{The peculiar case of $d=2$}  \label{degal2}

The Bragg glass is thus the prototype
of an elastic glass phase with internal periodicity.
In its physics $d=2$ plays a particular role because thermal fluctuations
become important. Also $d=2$ applies to a variety of experimental situations
such as magnetic bubbles, charge density waves, colloids, random XY model.
In addition, further analytical methods become available
\cite{cardy_desordre_rg,goldschmidt_dynamics_flux,goldschmidt_houghton_statics,toner_log_2}
to analyze the problem with and without dislocations.

When dislocations are {\it excluded} by hand a glass phase
is believed to exist for $T< T_g$ in $d=2$.
For the simpler $N=1$ component model
(i.e the random field XY model) this was shown by
Cardy and Ostlund\cite{cardy_desordre_rg} (CO) who used $n$ replicated
coupled XY models,
mapped them onto a Coulomb gas with $n(n-1)/2$ types of vector charges
and constructed the RG equations. They set $n \rightarrow 0$
implicitly assuming a replica symmetric fixed point.
The resulting RG equations, valid near $T_g$, possess a non-trivial
perturbative fixed point for $T<T_g$ at weak disorder
$g=g^* \propto T_g-T$.
CO concluded that a ``glass'' phase exists,
controled by this fixed point. In this phase,
one coupling constant flows
to infinity, a rather peculiar feature.
This $N=1$ model is known to describe
flux lines lying in a plane (a system where topological defects are indeed
excluded by construction).
These results and their connections to glass phases of flux lines
were extended in
\cite{toner_log_2,hwa_fisher_flux,nattermann_flux_creep,carpentier_ledou_triangco,carraro_nelson},
and the disorder averaged correlation function
$\tilde{B}(x)= \overline{\langle u(x)- u(0)\rangle ^2}$ was
found to grow as $\tilde{B}(x) \sim A {(\log|x|)}^2$, faster than
$\tilde{B}(x) \sim T \log|x|$ which holds
in the high temperature phase and for the pure system.

By analogy it was
argued \cite{giamarchi_vortex_long}
that similar results hold in the case of the triangular lattice.
This was verified explicitly recently
\cite{carpentier_ledou_triangco} using the RG
on the fully coupled $N=2$ component model
required to describe a triangular lattice.
The calculation leads to a glass phase for $T < T_g$
described by {\it a plane} of perturbative
fixed points. The growth of relative displacements is found to be asymptotically
isotropic with $u_T^2 \sim u_L^2 \sim A_1 \ln^2 r$, with
universal subdominant anisotropy $u_T^2 - u_L^2 \sim A_2 \ln r$.
where $A_1$ and $A_2$ depend continuously
on temperature and the Poisson ratio $\sigma$.
The fixed points obtained in \cite{cardy_desordre_rg} and in
\cite{carpentier_ledou_triangco} are thus
the natural continuation to $d=2$ of the non trivial
fixed point which describes the Bragg glass phase for $d=3$
(and $N=1$, $N=2$ respectively).
$d=2$ thus appears as the lower critical dimension of
this fixed point. While for $d \ge 2$ the Bragg glass fixed point is
a {\it zero temperature fixed point} where temperature
is (dangerously) irrelevant, in $d=2$ the glass phase is
described by a
{\it line of fixed points} where temperature is {\it marginal}.
Since entropy still plays a role in these $d=2$ periodic glasses
they can be called {\it marginal glasses}. This is illustrated
in their dynamics: the dynamical exponent $z$ (such that $t \sim x^z$)
was computed below $T_g$ (assuming equilibrium dynamics,
equivalent to assuming replica symmetry in the statics - see below)
for $N=1$\cite{goldschmidt_dynamics_flux,tsai_dynamics_flux}
and  for the triangular lattice\cite{carpentier_ledou_triangco}
$N=2$:
\begin{eqnarray}
z - 2 &\sim& 2 e^{\gamma} \tau \qquad\qquad {\rm CO\;\;model} \\
z - 2 &\sim& 3 e^{\gamma} \tau
\frac{ (2+ \alpha) (\frac{2-\alpha}{2+ \alpha})^{\frac{2-\alpha}{4}} }{
2 I_0(\alpha/2) - I_0(\alpha)} \qquad\qquad {\rm triangular\;\;lattice}
\end{eqnarray}
It is also continuously varying with the reduced temperature $\tau = (T_g-T)/T_g$
and the Poisson ratio $\sigma$ through $\alpha= 2 (1+\sigma)/(3-\sigma)$.
$\gamma$ is the Euler constant and $I_0$ the modified Bessel function).
This finite $z$ dynamical exponent yields the following $I$-$V$ (equivalently $v$-$f$)
characteristics:
\begin{eqnarray} \label{creep2d}
v \sim \mu_0 f \left(\frac{f}{f_c}\right)^{\frac{z-2}{2}}
\end{eqnarray}
Note that this is different from the creep law (\ref{creeplaw}) valid in
$d>2$ which is typical of a $T=0$ fixed point with a dynamical
exponent formally $z = \infty$. The $d=2$ result (\ref{creep2d})
can be interpreted as barriers which increase logarithmically
$U(j) \sim \ln (1/j)$ with decreasing current rather than algebraically as in $d>2$.

There are also some interesting issues related to the
possibility of replica symmetry breaking in this phase.
The Gaussian replica variational method (GVM)
described above, when applied to this model, leads to a one step replica
symmetry breaking (RSB) solution below $T_g$ and thus correctly predicts
the transition but yields mean squared relative displacements
growing as $u^2 \sim T_g \ln r$. This is a different result
from the replica symmetric (RS) RG prediction $u^2 \sim A_1 \ln^2 r$.
The GVM, being by construction an approximation, neglects some
non-linearities
and has no a priori reason to yield the exact result. However it
may be a hint that, if allowed, RSB will occur in this model.
It was indeed shown within the RG
\cite{ledoussal_rsb_prl,kierfeld_co_rsb} that the Cardy
Ostlund RS-RG flow
is unstable to an infinitesimal RSB perturbation at and below $T_g$.
The issue was thus raised \cite{ledoussal_rsb_prl} of whether
the RS-RG may miss some of the physics related to RSB.
The numerical studies presently available show
discrepancies \cite{batrouni_numerical_cardy,numRSGM2},
and their analysis is not yet fully satisfactory. Though there is
a more recent trend
\cite{marinari_sinegordon_numerics,lancaster_numerics_co,%
rieger_zerot_co,middleton_zerot_co}
towards a behavior consistent with the RSRG, it is
still only qualitative agreement.
Since a more careful treatment of the effects of RSB may
reveal that deviations from the RS-RG result are small
\cite{footnote_rsrg},
e.g only in the amplitude of the $\log^2 r$ \cite{ledoussal_rsb_prl},
more precise numerical tests should be performed.

When topological defects are {\it allowed} the above
picture will probably be modified. Indeed it was
shown in \cite{cardy_desordre_rg,goldschmidt_dynamics_flux} that
for the $N=1$ component model these defects are perturbatively
relevant near $T_g$. As argued
in \cite{giamarchi_vortex_long} $T_g$ for the triangular
lattice is well above the KTNHY melting temperature $T_m$ and
dislocations should then be relevant near $T_g$
for the $N=2$ triangular lattice as well.
At low temperature however, much less is known
about the importance of dislocations. The common
belief \cite{blatter_vortex_review},
which is by no means rigorously established, is that
if dislocations are allowed, no true glass phase will
exist at $T>0$ in $d=2$. In the
simpler random phase shift model
(which does capture some of the physics of
the full problem), a high temperature
phase with unbound dislocations
was found to be {\it reentrant} at low temperatures
in \cite{rubinstein_nelson_shraiman,nelson_elastic_disorder}
suggesting the importance of topological defects at
low temperature. It was pointed out in
\cite{giamarchi_vortex_long}, from a study of the
CO RG flow, that at low temperature the scale
at which the lattice is effectively dislocation-free
(i.e the distance between unpaired dislocations) can be
{\it much larger} than the translational length $R_a$. Thus
even in $d=2$ the Bragg glass fixed point may be
useful to describe the physics, as a very long crossover
or maybe directly at $T=0$. It was also pointed out
in \cite{giamarchi_vortex_long2} that the conventional
CO RG will not be adequate at low temperature since it
assumes a {\it thermalized description} of the vortices,
neglects important effects such as the
pinning of dislocations by disorder and the position dependence
of their fugacity.
A similar idea was recently proposed and pushed
further by Nattermann et al. \cite{natt1} who reconsidered the simpler
random phase shift model. They explicitly showed
that \cite{rubinstein_nelson_shraiman}
was incorrect at low temperature and proposed a
modified approach which leads to a phase which is
{\it defect free } at low temperature. Though these
new approaches need to be put on a firmer theoretical
footing (there are several underlying assumptions)
and though it is still an open question
how they carry to more complicated and
realistic elastic models, it is remarkable that
the results of \cite{natt1} do provide another
non trivial example, besides the $d=3$ Bragg glass,
of a case where topological defects are
{\it less relevant} than
is naively assumed.

Thus the question of dislocation is subtle even in $d=2$. Using
naively the RG one would conclude that there are always
dislocations. This is intuitively clear since in $d=2$ dislocations
are simply point like defects and thus much easier to create
by disorder. But it could also be too naive, since disorder
obviously modifies also the interactions between the defects
which are mediated by the elastic interactions. Thus
it is still an open question whether this is
really correct. In any case, even if it was,
the length between unpaired dislocations $R_D$ clearly grows
{\it much faster} than $R_a$ (it can be estimated
as $R_D \sim R_a e^{\ln^{1/2}(R_a)}$ see
\cite{giamarchi_vortex_long}). By reducing the disorder
and temperature one can thus, even in $d=2$, have
arbitrarily large dislocation free regions
where the main source of translational order decay
is from elastic deformations.
Regimes where $R_D \gg R_a$ have indeed been
seen in magnetic bubble experiments (see discussion in
\cite{giamarchi_vortex_long}).

Let us conclude this section by noting the amusing
twist by which the CO model, on which the proposal
of the ``vortex glass'' phase \cite{fisher_vortexglass_short}
was based originally,
has turned out, upon further
analysis, to be of a totally different nature.
Indeed the CO glass phase is rather the continuation
to $d=2$ of the topologically ordered
Bragg glass phase which we have argued exists
as a thermodynamic stable phase in $d=3$.

\section{Dynamics of driven disordered lattices} \label{dynsec}

Obtaining a quantitative
description of the dynamics of driven interacting systems
with disorder is the next challenging problem.
Beyond vortex lattices it
is also important for several other experimental systems
such as Wigner crystals \cite{andrei_wigner_2d} moving under an applied
voltage,
lattices of magnetic bubbles \cite{seshadri_bubbles_long} moving under
an applied magnetic field gradient, Charge Density Waves (CDW)
\cite{gruner_revue_cdw} and colloids \cite{murray_colloid_prb}
submitted to an electric field,
driven Josephson junction arrays. For many of these systems transport
measurements are a useful way to probe the physics of
the system, and sometimes the only way if direct imaging cannot be
performed. Dynamic properties have thus been studied for some time,
especially for the case of CDW or for driven manifolds and their relation to
growth processes\cite{krim_growth_review} described by the Kardar Parisi
Zhang (KPZ) equation\cite{kardar_parisi_zhang,kardar_review_lines},
using a variety of
methods that we will only briefly review here. Curiously the similar problem
of a periodic {\it lattice} (with additional periodicity transverse to
the direction of motion) was not scrutinized until very recently,  maybe because it was
naively thought that it falls in the same class as the above problems.
Fortunately, as for the statics, (transverse) periodicity drives
again the system to a novel behavior, the richness of which is far
from being understood. We will thus mainly devote the rest of this
section to this particular case of dynamical problems.

To tackle the dynamics of such periodic systems, it is important to
know whether topological defects in the structure
are generated by disorder, temperature and the driving force
or their combined effect. Indeed if such defects exist, the flow will
not be elastic, but will turn into the so-called plastic flow, with a
radically different behavior. It is an important and still largely open
question to determine
when plastic rather than elastic motion occurs but quite generally
one expects plastic motion for strong disorder situations, at high
temperature, and probably close to the threshold in low dimension (for
CDW see e.g.\cite{coppersmith_defects_cdw}). This is confirmed by
experiments and numerical simulations. Indeed,
there has been a large number of studies on
plastic flow \cite{shi_berlinsky},
e.g in the context of superconductors where
a $H$-$T$ phase diagram
with elastic flow regions and
plastic flow was observed\cite{higgins_second_peak}.
Several experimental new effects have been attributed to it
such as the peak effect \cite{bhattacharya_peak_effect1,higgins_second_peak},
unusual broadband noise \cite{marley_broadband_noise} and
fingerprint phenomena in the I-V curve
\cite{bhattacharya_fingerprints,hellerqvist_ordering_short,hellerqvist_ordering_long}.
Close to the threshold and in strong disorder situations
the depinning is known to proceed \cite{gronbech_jensen_filamentary} through
filamentary flow in what can be called ``plastic channels''
\cite{jensen_plasticflow_short,jensen_plasticflow_long}
between pinned regions. Despite  numerous
studies, mostly numerical ones, a detailed theoretical understanding
\cite{watson_fisher_plastic} of this regime is still sketchy.

One could expect to be in a better position to attack the problem
of the elastic flow. At first the task seems formidable.
Experience from other glassy
systems, such as spin glasses, has taught us to expect an extremely
complicated dynamics due to the peculiar features of the energy
landscapes \cite{cugliandolo_dynamics_review}.
Generally three main dynamical regimes can be established.
Far below the threshold the system can move only through thermal
activation. This is the creep regime where qualitative arguments
have been developed \cite{feigelman_collective}.
One would like to check whether these
rather phenomenological
arguments can be confirmed by more direct (and hopefully rigorous)
dynamical calculations. The second regime, near the elastic depinning
transition, has been intensely investigated in similarity with standard
critical phenomena (see e.g
\cite{fisher_depinning_meanfield,narayan_fisher_depinning,%
nattermann_stepanow_depinning})
where the velocity
plays the role of an order parameter.
The third regime, which is the one we will
concentrate on here, is far above the threshold.
In this regime things may look more rosy,
since one could also imagine that a sliding system averages
in fact enough over disorder, to recover a simple enough behavior, in
fact much simpler than in the statics. Indeed it was observed
experimentally some time
ago in neutron diffraction experiments \cite{thorel_neutrons_vortex},
and in more details recently \cite{yaron_neutrons_vortex}
that at large velocity the
vortex lattice is more translationally ordered than at
low velocity. In this regime, since the velocity is large, one is not so
much interested in the $v$-$F$ characteristic, but much more on the
positional
properties of the moving system. A question of prime interest
is thus whether at large enough velocity glassy effects disappear and
whether one recovers a perfect lattice.

Before concentrating on this issue and seeing that the answer
crucially depends on the periodic nature of the driven system,
let us look at the general behavior of driven lattices and first
establish the equation of motion.

\subsection{The basic equation}

The conventional description for the dynamics of these systems
is in terms of overdamped dynamics with a microscopic friction
coefficient $\eta$. Let us denote by $R_i(t)$ the true position of an individual
particle in the laboratory frame and assume that the lattice as a whole
moves with a velocity $v$.  We thus introduce the displacements
$R_i(t)=R_i^0+v t+u_i(t)$
where the $R_i^0$ denote the equilibrium positions in
the perfect lattice with no disorder. $u_i$ represent the displacements
compared to a moving perfect lattice (and corresponds to the
position of the i-th particle in the moving frame). The definition of
$v$ imposes $\sum_i u_i(t) = 0$ at all times. For a manifold $i$
would be a continuous index spanning the internal dimension of the
manifold. Using these variables the equation of motion can be written
in the {\it laboratory frame}:
\begin{eqnarray} \label{labasem}
\eta \frac{d u_i(t)}{dt} = - \frac{\delta H_{el}}{\delta u_i}
+ \int_r \partial V(r) \delta(r - R_i^0+vt+u_i(t) ) +
f - \eta v + \zeta(R_i(t),t)
\end{eqnarray}
where $f$ the external uniform
force driving the system and $\zeta$ is a thermal noise which
satisfies $\overline{\zeta(r,t) \zeta(r',t)}
= 2 T \eta \delta(r-r') \delta(t-t')$.
The other two forces acting on
the system are the elastic force, derivative of the elastic Hamiltonian
$H_{\rm el}$ (\ref{real}), and the pinning force, coming from the
coupling (\ref{complic}) to the random potential. As for the statics it
is fruitful to take the continuous limit. Using the decomposition of the density
(\ref{transparent}) allows to rewrite (\ref{labasem}) in the simple form
\begin{equation} \label{eqmotion}
\eta \partial_t u_{rt}^\alpha + \eta v \cdot \nabla u_{r t}^\alpha =
\int_{r'} \Phi_{\alpha \beta}(r-r')  u_{r' t}^\beta
+ F_{{\rm pin}}^{\alpha}(r,t)
+ f_{\alpha}-\eta v_{\alpha} + \zeta_{\alpha}
\end{equation}
where $\Phi_{\alpha \beta}(r-r')$ is the elastic matrix. The convection term
$\eta v \cdot \nabla u_\alpha$ comes from the standard Euler
representation when expressing the displacement field in the laboratory
frame. Note that this term is {\it not} the gradient of a potential,
as a consequence of the fact that the system is out of equilibrium
with energy constantly injected and dissipated. The pinning force
is given by:
\begin{equation} \label{pinham}
F^{{\rm pin}}_{\alpha}(r,t) = - \delta {H_{{\rm pin}}}/\delta
u_{\alpha}(r,t) =  V(r)\rho_0\sum_{K} iK_\alpha \exp
(iK\cdot(r-vt-u(r,t)))
\end{equation}
(as for the statics we only write the important Fourier components).
 A manifold, lacking the periodicity, can also be described by
(\ref{pinham}) simply by letting the discrete sum over the reciprocal
lattice vectors $K$ become an integral $\int dK$ to reproduce the
$\delta$ function of the density on the manifold.

(\ref{eqmotion}-\ref{pinham}) is the complete equation one would have to
solve, and again one is faced with the nonlinearities in
(\ref{pinham}): they both prevent one from solving the
equation and of course lead to most of the interesting effects.

\subsection{Critical force and large $v$ expansion} \label{largev}

To tackle these formidable equations (\ref{eqmotion}-\ref{pinham}),
the first angle of attack, again pioneered by Larkin \cite{larkin_largev},
and by Schmidt and Hauger \cite{schmidt_hauger},
is to perform a large velocity expansion of (\ref{eqmotion}).
Indeed at large $v$, (\ref{pinham}) oscillates rapidly due to the terms
in $Kv t$ and
vanishes\footnote{At that point the astute reader will have noticed some
forthcoming problems from the modes such that $K\cdot v=0$. We will come
back to that point later.}. One can then compute the
displacements $u$ in an expansion in $1/v$.
Solving at first order and using the corresponding
expression of $u$, (\ref{pinham}) gives a correction to the average
velocity \cite{larkin_largev,schmidt_hauger}
\begin{eqnarray} \label{hauger}
\delta v_\alpha = - \frac{1}{2} \sum_K  \sum_{I=L,T}
\int_{BZ}  \frac{dq}{(2 \pi)^d}
K_\alpha (K.P^I(q).K) \Delta_{K} \frac{v.(K+q)}{\Phi^I(q)^2 + (\eta
v.(K+q))^2}
\end{eqnarray}
where $P^{L,T}(q)$ are the standard longitudinal and transverse
projectors and $\Phi^{L,T}(q)$ the elastic energy of longitudinal
and transverse modes, e.g in $d=2$ at small $q$, $\Phi^{L}(q)=c_{11}q^2$ and
$\Phi^{L}(q)=c_{66} q^2$ where $c_{11}$, $c_{66}$ are respectively the
bulk and shear modulii.

One can push this formula, which is valid only at large $v$, beyond its
domain of validity to estimate\cite{blatter_vortex_review}
in a very qualitative way the value of the
threshold field by the criterion $\delta v/v \sim 1$ replacing $v$ by
$f_c/\eta$. More interestingly, in addition to giving an estimate for
$f_c$ which is found to be consistent with the Larkin Ovchinikov
arguments of section~\ref{simplesec}, the large $v$ expansion
allows in principle to compute the displacement correlation function.
This was done in \cite{koshelev_dynamics} where it was
concluded that at low $T$ and above a certain velocity
the moving lattice becomes a perfect crystal again at an effective
temperature $T^{\prime }=T+T_{sh}$.
The effect of pinning was then described \cite{koshelev_dynamics}
by some effective {\it shaking temperature}
$T_{sh}\sim 1/v^2$ defined by the relation
$\langle |u(q)|^2 \rangle = T_{sh}/c_{66} q^2$.
The physical picture that emerges from the naive large $v$ expansion
seems at first very reasonable since the system in fast motion
averages enough over disorder. Since the disorder vanishes and is
replaced by ``thermal'' effects, this approach would suggest bounded
displacements in $d>2$ and absence of glassy properties in the moving
solid. At least at large enough velocities, the dynamics would thus
seem much more simple than the corresponding statics !

However, the problem is more complicated than it looks and this
naive approach is incorrect. For reasons that we will explain in
section~\ref{themov}, the large $v$ expansion is invalid
\cite{giamarchi_moving_prl,giamarchi_mglass_long} and
transverse periodicity leads to (well hidden) divergences in
perturbation theory. In addition, due to the driving of the system,
other relevant terms, such as random forces are generated.
As for the statics, a correct study of the problem requires to fully
treat the nonlinearities of (\ref{pinham}), and sharper tools than
mere perturbation expansion are needed.

\subsection{Methods and what follows} \label{generated}

Fortunately, more powerful methods exist. The most standard
one is to introduce a field
theoretical description of (\ref{eqmotion}) which we write in a compact
form:
\begin{equation} \label{startexp}
{(R^{-1})}^{\alpha \beta}_{rt r't'} u^{\beta}_{r't'} = f_{\alpha} -
 \eta v_{\alpha} + f_{\alpha} (r,t,u_{rt})
\end{equation}
To do so, one introduces the Martin-Siggia-Rose-de Dominicis-Janssen
generating functional
\cite{martin_siggia_rose} given by
\begin{equation}
Z[h,\hat{h}] = \int Du D\hat{u} e^{- S[u,\hat{u}] + \hat{h} u + i h \hat{u}}
\end{equation}
where $\hat{h},h$ are source fields. The MSR action corresponding to the
equation of motion (\ref{startexp}) is
\begin{equation} \label{totalmsg}
S[u,\hat{u}]  =   S_0[u,\hat{u}] + S_{int}[u,\hat{u}]
\end{equation}
with
\begin{eqnarray}
S_0[u,\hat{u}] & = & \int_{rtr't'} ~ i \hat{u}^{\alpha}_{rt}
(R^{-1})^{\alpha \beta}_{rt,r't'}  u^{\beta}_{r't'}
- i \hat{u}^{\alpha} ( f_{\alpha} - \eta_{\alpha \beta} v_{\beta} )
- \eta T \int_{r,t} (i \hat{u}^{\alpha}_{rt}) (i \hat{u}^{\alpha}_{rt})
\label{freemsg} \\
S_{int}[u,\hat{u}]  & = & - \frac{1}{2} \int dr dt dt' (i \hat{u}^{\alpha}_{rt})
(i \hat{u}^{\beta}_{r t'})
\Delta^{\alpha \beta}(u_{rt} - u_{rt'} + v (t - t'))
\label{disordermsg}
\end{eqnarray}
where we recall that $\Delta(u-u')$ is the disorder correlator.
Causality imposes that $R_{rt,r't'}=0$ for $t'>t$
and the Ito prescription for time discretization
implies $R_{rt,r't}=0$. In such driven problems, space symmetry is
broken by the motion and $C_{-r,t}
\neq C_{r,t}$ when $v$ is non zero.

This formalism is widely used to study dynamical problems.
It has the advantage of treating separately the correlation
$C^{\alpha,\beta}_{rt,r't'} = \overline{ \langle u^{\alpha}_{rt} u^{\beta}_{r't'} \rangle }$
and response functions
$R^{\alpha,\beta}_{rt,r't'} = \delta \overline{ \langle u^{\alpha}_{rt} \rangle }/
\delta h^{\beta}_{r't'}$ which measures the linear response to a perturbation
applied at a previous time. They are obtained from the
above functional as
$C^{\alpha \beta}_{rt,r't'} = \langle u^{\alpha}_{rt} u^{\beta}_{r't'} \rangle_S $
and $R^{\alpha \beta}_{rt,r't'} = \langle u^{\alpha}_{rt} i \hat{u}^{\beta}_{r't'} \rangle_S$
respectively. Although these two functions are usually related by the
fluctuation dissipation theorem $ T R^{\alpha \beta}_{r,t}
= - \theta(t) \partial_t C^{\alpha \beta}_{r,t}$ for equilibrium
problem, this does not need to be the case in dynamical ones. The MSR
formalism is thus able to tackle out of equilibrium dynamics
\cite{cugliandolo_ledoussal_manifold,cugliandolo_ledoussal_zerod}
for which the fluctuation dissipation theorem (FDT)
does {\it not} hold. This is the case here
we are studying a moving system which does not
derive from a Hamiltonian.

Starting from MSR one can either use a dynamical mean-field theory
\cite{cugliandolo_pspin,cugliandolo_ledoussal_manifold}, or
since MSR is a field theoretical formulation
derive renormalization group equations by integrating over short scales.
Using such a renormalization group one can go beyond the large $v$ expansion
and access properties at the depinning transition. Such a procedure was
pioneered to study the depinning of manifolds in
\cite{nattermann_stepanow_depinning,narayan_depinning_short,%
narayan_fisher_depinning}.
A similar situation to that of the static functional renormalization
develops. The disorder correlator $\Delta(u)$
flows to a fixed point function (which corresponds to
the threshold fixed point $v=0$) which is {\it non analytic}. This non analyticity
was shown to be directly related to the critical force by
$f_c \sim \Delta'(0^{+})$. As can be expected from Larkin's model,
the scale at which the non analyticity appears
is the Larkin length $R_c$. This method suitably generalized
\cite{giamarchi_m2s97_vortex,giamarchi_mglass_long} allows also for the
study of the moving periodic system.

Within the MSR formalism one can also study the generation
of different additional terms in the equation of motion. In particular,
as was first observed in growth processes, and later in driven
manifolds\cite{kardar_parisi_zhang,kardar_review_lines},
the application of an external driving force
breaks the symmetry $u_x \to -u_x$
(as can be seen on (\ref{startexp})). This allows, from pure
symmetry considerations, the generation of non linear terms, such as
$(\nabla u)^2$. These terms, the so called KPZ terms, may be relevant
and change the large scale behavior drastically. The effect of such
terms in driven dynamics is a subject of active investigations
\cite{kardar_review_lines,chen_kpz_dynamics}.
One can show explicitly that in the
present problem, not only these KPZ like terms, but also linear terms
which are allowed by symmetry, are indeed generated
\cite{giamarchi_mglass_long}
from a finite cutoff effect. Since their bare value is very small
they may be important only at very large length scales.
Other non trivial terms can be generated in such dynamical
systems. One of the simplest is the random force. It is
intrinsically
a non equilibrium effect since it cannot exist in the
statics by the symmetry $u\to u+a$. Usually the
existence of such terms are conjectured \cite{balents_dynamics_vortex}
since their explicit
calculation from the original equation of motion (\ref{eqmotion}) is
difficult. We will come back to that point and give an explicit
derivation of this term
in
the next section.

\subsection{Periodicity peculiarities: the Moving glass} \label{themov}

We now come back to the detailed physical properties of the
periodic structures. The physics becomes
transparent when looking at the pinning force (\ref{pinham}):
all the modes of the disorder such that $K \cdot v = 0$ do {\it
not} have a direct time dependence (some time dependence can be
introduced through the displacement field $u$), and will {\it not}
average out even at large velocity. This is assuming that motion occurs
along one of the lattice direction, which is physically reasonable.
In such case the large $v$ expansion
breaks down ! The equation of motion thus contains a static disorder
component perpendicular to the direction of motion.
\begin{equation}
F_\alpha ^{{\rm stat}}(r,u) = V(r)\rho_0\sum_{K.v=0} iK_\alpha \exp
(iK\cdot(r-u))
\end{equation}
Since this force is perpendicular to the velocity, it has no impact on
the velocity correction itself, for which the large $v$ expansion is
still perfectly valid. On the other hand the correlation functions of the
displacements will
be drastically affected.

Again the periodic system proves its
peculiar nature. Note that here periodicity {\it perpendicular} to the
direction of motion is the crucial ingredient. The effects described here
will thus be absent for systems for which such transverse periodicity
does not exist such as single $q$ CDW (for which the density is only
modulated in one direction\cite{gruner_revue_cdw}), or manifolds pulled
orthogonally to their
direction, such as in growth phenomena. On the other hand systems
such as vortex systems or the Wigner crystal are directly affected.

\subsubsection{Beads on a string}

The resulting physics can be easily understood by focusing on the
component
of the displacement $u$ affected by this remaining static disorder.
Although this is an approximation it provides a clear enough view of the
physics and can be confirmed by a more rigorous renormalization
calculation\cite{giamarchi_mglass_long}. Clearly only transverse
displacements are affected, and the static equation is
\begin{equation} \label{lastatic}
\eta \partial_t u_{rt}^y + \eta v \cdot \nabla_x u_{r t}^y =
c \nabla^2 u_{r t}^y
+  V(r)\rho_0\sum_{K.v=0} iK_y \exp
(iK\cdot(r-u))
\end{equation}
where we denote generically by $y$ the transverse directions
and we have chosen isotropic elasticity for simplicity.
If it were not for the linear term $v \nabla_x u$, this equation would
be identical to the one describing the relaxational dynamics of
a periodic manifold in a random potential (without driving).
The linear term
is the one taking the dissipation into account and making the problem
different from an Hamiltonian one. From the point of view of the
solution however this term merely introduces a different scaling between
the direction of motion ($x$) and the perpendicular directions ($y$),
$L_x \sim L_y^2$. Due to the presence of the static disorder, one
thus expects the transverse components to present all the characteristics
of a static disordered elastic system and thus to exhibit pinning, to
have unbounded growth of displacements etc. In short to be a glass.
This time the periodicity makes the system more disordered than was
naively expected !

Of course there are various important differences with the static problem.
The simplest one comes from the scaling $L_x \sim L_y^2$, which makes
the disorder in the moving system only relevant for $d \leq 3$.
For $d > 3$ the moving system
is not a glass but a perfect crystal at weak
enough disorder or large velocities.
Apart from this rescaling the physical properties of the
moving system presents some similarities with the one of the elastic disordered
systems exposed in section~\ref{dissec}. In particular there will
be a {\it static} solution for the transverse displacement, becoming
rough at a length scale $Rx \sim R_y^2$, analogous to the $R_a$ of the
static problem. Estimates a la Fukuyama-Lee similar to the one of
section~\ref{charlength} give:
\begin{equation} \label{fukulee2}
R_y^a \sim (a^2 v c/\Delta)^{1/(3-d)}, \qquad
R_x^a = v (R_y^a)^2/c
\end{equation}
The moving glass is highly anisotropic since $R_x^a/R_y^a$
diverges as $v\to \infty $.

\begin{figure}
\centerline{\epsfig{file=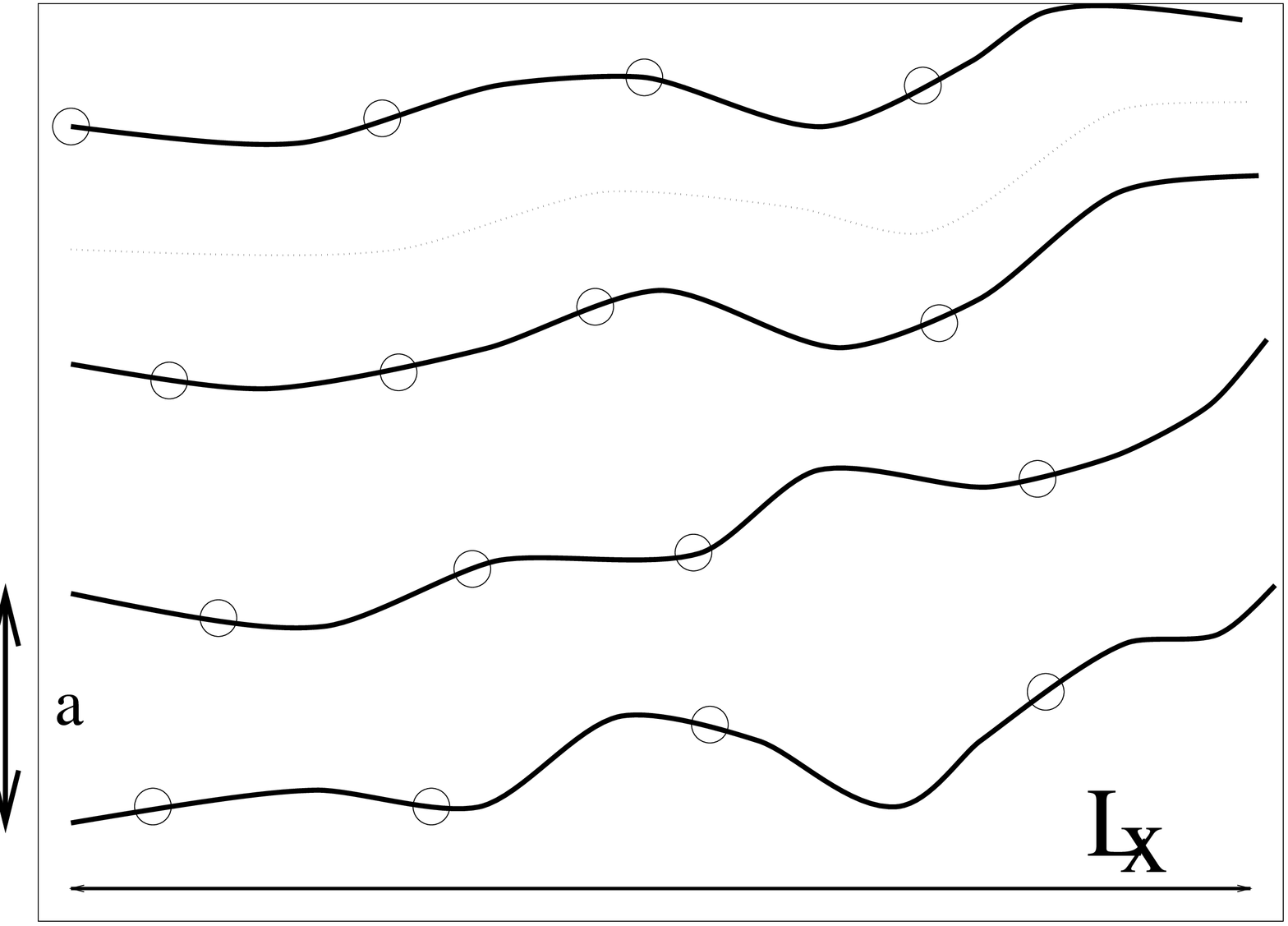,angle=-90,width=7cm}
            \epsfig{file=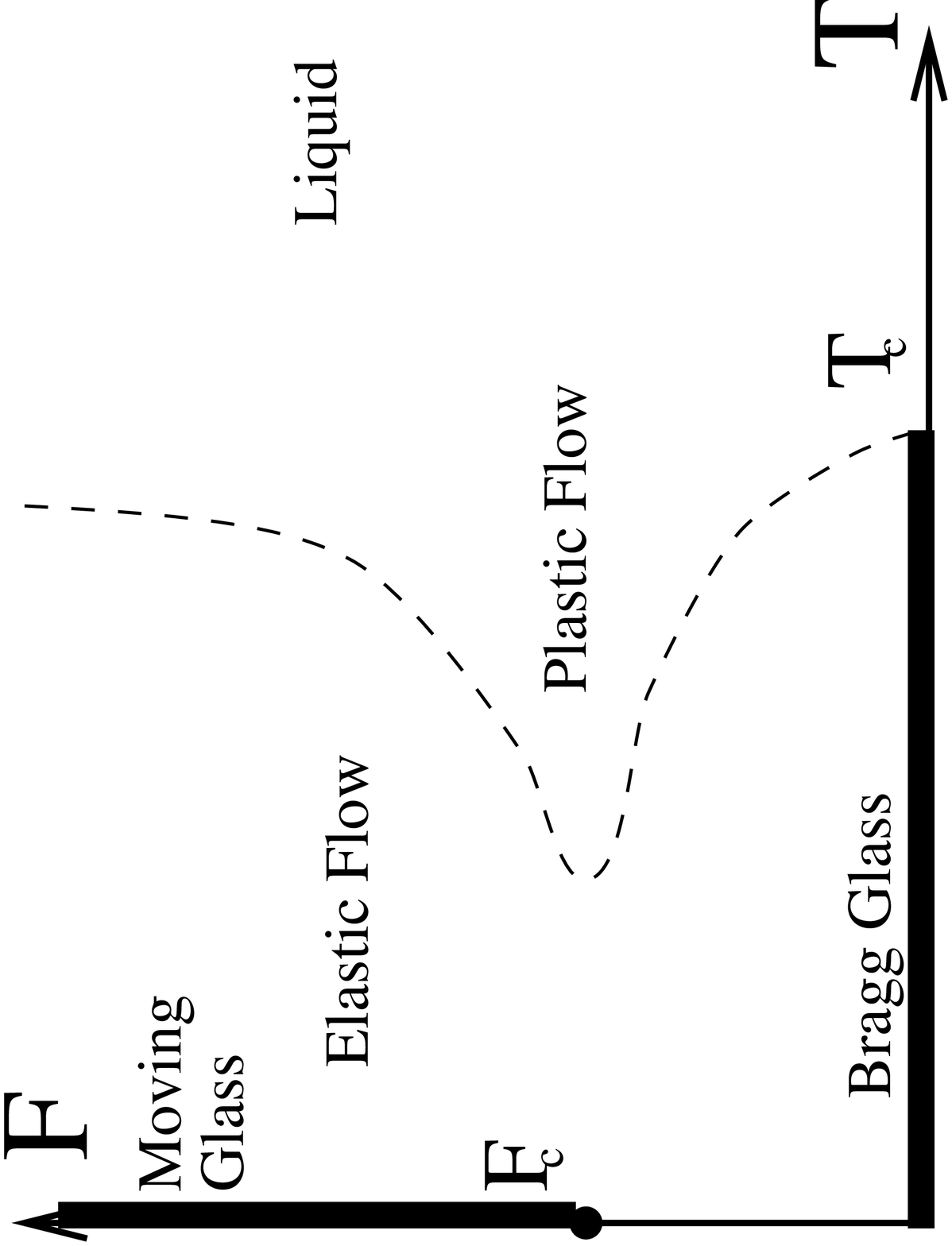,angle=-90,width=7cm}}
\caption{a) Motion in the moving glass occurs through static channels
wandering at distance $a$ over lengths $L_x \sim L_y^2$. If
dislocations are present ($d=2$ or strong disorder in $d=3$) they
should lead to a decoupling of channels, as indicated by the dotted
line. A square lattice was represented for simplicity.
b) Phase diagram in force $F$, temperature $T$, for weak disorder
and in $d=3$. At zero
external force the system is in the free Bragg glass state. At large
velocities, in the moving Bragg glass one. This suggests that in this
case the depinning transition could be purely elastic.
\label{figchan} }
\end{figure}

Thus, the moving vortex configurations
can be described in terms of {\it static channels}
that are the easiest paths where particles follow
each other in their motion like beads on a
string\cite{giamarchi_moving_prl,giamarchi_mglass_long}.
Channels in the elastic flow regime are fundamentally different in nature
from the one introduced to describe slow plastic motion between
pinned islands \cite{jensen_plasticflow_short,jensen_plasticflow_long}.
In the moving glass
they form a manifold
of elastically coupled, almost parallel lines or sheets (for
vortex lines in $d=3$) directed along $x$
and characterized by some transverse wandering $u_y$.
In the laboratory frame they are determined by the static disorder
and do not fluctuate with time.
In the moving frame, since each
particle is tied to a given channel which is now
moving, it indeed wiggles and dissipates
but the motion is highly correlated with the neighbors.
An image of this channel picture is shown in figure~\ref{figchan}.
Such channels were subsequently observed in numerical simulations
\cite{moon_moving_numerics} and in recent decoration in motion
experiments \cite{marchevsky_decoration_channels}.
The physical picture emerging is thus completely different
from the image of a perfect lattice submitted to an extra
shaking temperature. This concept may be correct however in
the liquid.

Determining the roughness of the channels is not an easy task. It is of
course impossible to directly borrow the static result since the problem
is by nature dynamics because of the dissipation term $v \nabla_x u$,
and as we saw in section~\ref{generated}, many terms can be generated
due to the motion. The full solution of the problem is not known, and in
particular the effects of the nonlinear KPZ-like terms. If one ignores
such terms, it is possible to apply the FRG to extract
the roughness of the channels.
We quote here only the RG equation for the disorder term, for the
periodic structure at $T=0$ (for the complete calculation see
\cite{giamarchi_mglass_long})
\begin{equation} \label{shortequ}
\frac{ d \Delta(u)}{dl}
= \Delta(u) +  \Delta''(u) ( \Delta(0)  - \Delta(u) )
\end{equation}
where a factor $\frac{1}{4 \pi v c} \epsilon$, with $\epsilon=3-d$
has been absorbed in $\Delta(u)$ (chosen to be of period $1$).
For $d>3$ disorder renormalizes to zero and the moving system is a
crystal. For $d < 3$ $\Delta$ flows to a new fixed point
$\Delta^{*}(u) = \Delta(0)(l) + u^2/2 - u/2$,
showing that the static disorder is still relevant in the moving
structure
(with the same conclusion\cite{giamarchi_mglass_long} in $d=3$).
This new fixed point describes the moving glass phase at $T=0$.
The value of $\Delta(0)(l)$ grows unboundedly as
$\Delta(0)(l) = \Delta(0) e^{\epsilon l}$ which indicates the existence
of a random force along the $y$ direction, generated under
renormalization. A similar force is generated along $x$
\cite{giamarchi_mglass_long}. Thanks to the RG formulation we are able
to {\it explicitly} compute this random force
\begin{eqnarray}
\delta \Delta_0^{\alpha \beta} =
\sum_K \int_{q} K^4 K_{\alpha} K_{\beta} \Delta^2_K
\frac{ \eta^2 v^2 (K_x + q_x)^2 }{
(c^2 q^4 + \eta^2 v^2 (K_x + q_x)^2)^2}
\end{eqnarray}
where $\Delta_0$ is the correlator of the random force.
The divergences of $\Delta(0)$ does not spoil the above fixed point,
since one can always
separate the random force $\Delta(0)$ and  the non linear part
$\Delta(u) - \Delta(0)$.
The generated random forces will have very different impact
depending on the dimension. In $d=3$ displacements only grow logarithmically,
so the MG conserves quasi-long range translational order.
Thus similarly to the statics, the MG in $d=3$ at
weak disorder or large velocity is expected to retain perfect
topological order. In that case one would go from a static Bragg glass
without dislocations to a moving Bragg glass also without dislocations
(at large velocities). It is thus possible that in $d=3$ the depinning
occurs without an intermediate plastic region, leading to the phase
diagram of figure~\ref{figchan}.

In $d=2$ however displacements grow algebraically and dislocations are
more likely to appear. The existence of channels \cite{giamarchi_moving_prl}
then {\it naturally} suggests
a scenario by which dislocations affect the MG: when
the periodicity along $x$ is retained, e.g., presumably in
$d=3$ at weak disorder, the channels are coupled along $x$.
Upon increasing disorder or decreasing velocity in $d=3$,
or in $d=2$, decoupling between channels can occur,
reminiscent of static decoupling in a layered geometry
\cite{carpentier_bglass_layered} (see
also \cite{balents_mglass_comment_reply}).
Dislocations are then inserted between the layers,
naturally leading to a flowing smectic glassy state,
recently observed in $d=2$ numerical simulations
\cite{moon_moving_numerics}.
Indeed, the transverse smectic order is likely to be more
stable than topological order along x, because of particle
conservation \cite{giamarchi_mglass_long}.

The problem of the behavior of dislocations in the moving glass system
is of course still open, and constitutes as for the statics one of the
most important issues to understand.

\subsubsection{Transverse pinning}

As an important consequence of the existence of the MG,
barriers for transverse motion exist
once the pattern of channels is established. Thus the
response to an additional small transverse force $F_y$
is very non linear with activated behavior and hysteretic
behavior (history dependence). At $T=0$
and neglecting the dynamic part of the disorder a true
transverse critical current $J_y^c$ exists.
This can be seen by adding a transverse force in (\ref{eqmotion})
$J_y^c$ can then be estimated by balancing the pinning energy
with the transverse Lorentz force acting on a Larkin domain:
\begin{displaymath}
J_y^c=\frac{c}{\phi_0 r_f}\Delta^{1/2} (R_y^c)^{-(d-1)/2}(R_x^c)^{-1/2} \sim
 \widetilde{\Delta }
^{2/(3-d)}
\end{displaymath}
where $\tilde{\Delta }=\Delta /v$ is
an effective velocity-dependent disorder.
A more rigorous derivation can be obtained from the RG equation
(\ref{shortequ}). Since its fixed point has
a {\it non analyticity} at $u=0$, (leading to
$\Delta'(0^+) = 1/2$) there is a critical force,
determined at the Larkin length $L_y$:
\begin{equation} \label{critforce}
F_c = \int dq G(q) \Delta'(0^+) =
\frac{\epsilon}{4 \pi v c_y}
\sim (R^c_y)^{-2}
\end{equation}
(\ref{critforce}) coincides with the above more qualitative derivation.
In $d=3$ one finds:
\begin{equation}
R_c^y \sim e^{\frac{4 \pi v c_y}{\Delta_0}}
\end{equation}
The MG is thus dominated by the
competition between the random force and the critical force.

The transverse critical force is a subtle effect since
it apparently breaks the rotational symmetry of the problem.
In fact it is a purely dynamical effect due to barriers
preventing the system to reorient the channels in the direction
of the total applied force. It does {\it not} exist for a single driven vortex line
(or for any manifold driven perpendicular to itself) in a random potential,
except if the potential is sufficiently correlated in the direction of motion
(such as a periodic potential).
In some sense here the elasticity of the manifold provides the necessary correlations.

After the prediction of the moving glass,
the transverse critical force at $T=0$ in $d=2$ was observed in numerical
simulations\cite{moon_moving_numerics,ryu_mglass_numerics}
(see also \cite{koshelev_privcomm,holmlund_mglass_numerics})
and found to be a fraction of the longitudinal critical force.
These predictions
can be also tested in experiments on the vortex lattice, or other systems
such as colloids, magnetic bubbles or CDW, or
or numerical simulations. Additional physical consequences and
references can be found in
\cite{giamarchi_moving_prl,giamarchi_mglass_long}.

\subsection{Dissipative glasses}

The physics of a periodic moving system has thus some novel properties
and in particular was shown to preserve glassy effects (at least for $T=0$)
even in an out of equilibrium
regime where the dynamics is {\it non potential}.
We have proposed this physical system as a first example of
a ``dissipative glass'', i.e. a glass with a constant
dissipation rate in the stationary state. It seems to hint that
non potential dynamics can indeed exhibit glassy properties,
a question which one can ask in a more general context.

It was asked recently \cite{cugliandolo_nonpotential_prl}
in the context of spin systems. There it was also found,
within mean field, that some glassy properties survive, a conclusion
which was going against previous conclusions in mean field models.
In finite dimensional system (finite $N$)
we expect these non potential glassy effects to be even stronger
and even in some cases survive at finite temperature
\cite{giamarchi_mglass_long,ledou_wiese_ranflow}. With
hindsight one could in fact consider that a prototype of these
systems, even if oversimplified, is the example of a particle
diffusing in a random non potential flow with long range correlations,
a problem studied a long time ago \cite{ledou_diffusion_particle}
using a RG approach (it was also solved recently in the large $N$
limit \cite{ledou_polymer_longrange}).
Remarkably this problem already exhibits glassy effects
and a finite temperature fixed point. This prototype model,
and more generally all non potential problems
(including e.g the moving glass) are described by
a Fokker Planck operator whose spectrum is not
necessarily real (by contrast with potential problems
which are purely relaxational).
We also want to point out the
deep and interesting connection \cite{giamarchi_mglass_long}
that exists between the new type of dissipative glassy problems
described here and
the study of general {\it non hermitian} random operators.
These operators, which in their low dimensional versions
can be termed {\it non hermitian quantum mechanics},
appear in several problems recently studied
with a renewed interest (such as vortex lines with tilted
columnar defects \cite{hwa_splay_prl,chen_kpz_dynamics},
spin relaxation in random magnetic fields \cite{mitra_pld_rmn}
or again diffusion of particles and polymers in random flows
\cite{chalker_diffusion,ledou_diffusion_particle,%
ledou_wiese_ranflow}). Exploring this
connection further, as well as the question of
the classification of these glasses and the
study of their physical properties is still a largely open but extremely
challenging field.

\section{Conclusion} \label{concsec}

We have examined some aspects of the statics and dynamics of
disordered elastic systems in presence of disorder, both for manifolds
and periodic systems. Although these two classes of systems share many
basic properties, periodic structures
exhibit a new type of physics quite different and unexpected
compared to elastic manifolds. Indeed, for static properties,
periodicity helps the system to resist to the disorder and to preserve
quasi long range order, while still having the energy landscape
and many metastable states of a glass. In turn this preservation of
quasi-translational order leaves the system quite stable to the
proliferation of topological defects such as dislocations, much in the
spirit of an honest solid. This two-faced state (the Bragg glass) both
a lattice and a glass seems to be realized in vortex
systems, and to explain many of the observed experimental features.
The mere existence of the Bragg glass, having {\it no} unbound
dislocations at equilibrium, implies the existence of a transition upon
increasing the disorder (or equivalently the magnetic field). This
suggests that two glass phases might exist in vortex systems, that
could be distinguished by the presence or absence of free topological
defects. The existence of such a transition and of the Bragg glass
cannot be anticipated by looking at the physics of manifold alone, for
which disorder always wins leading to rough and defective ground states.
For the dynamics, periodicity plays again an important role,
but this time with quite
opposite effects. The manifold driven at high velocities offers no
surprises and would be quite ordered (though with some
anomalous KPZ type roughness) whereas the periodic system
remains a glass, with surprising properties such as the existence of a
transverse critical force.

The physics of disordered periodic systems offers thus a field rich of
prospectives and challenging problems. Of course both for the
statics and dynamics, the issue of topological defects is of paramount
importance. In the statics it is the key to understand the transitions
leading to the destruction of the Bragg glass, as well as the mysterious
strong disorder glass. The question is also particularly important for
two dimensional systems. For the dynamics the very question of the
presence of topological defects in the moving glass phase is at stake.
Solving this issue starting from large velocity is already a formidable task,
but could help us to understand what happens close to the threshold. Indeed
here again only simple cases, inspired from the manifold or CDW with
scalar displacements and no transverse periodicity, have been
considered previously.
As in the statics it is possible that the physics is modified in a quite
surprising way, and certainly all the issues about critical behavior
close to threshold, dynamics reordering, elastic to
plastic motion transitions, will have to be reconsidered. These
issues are of major theoretical concern but also of large
practical importance.

Another issue which is raised is how to handle the whole complexity
of all the new terms which are generated in the driven dynamics
of a lattice (in order to go beyond the simpler models which have
been analyzed here). Among such new terms are the random force term,
additional linear and non linear terms which are allowed by symmetry,
and their interplay with the quenched disorder terms which remains
due to transverse periodicity. This will require to go further than
the whole topics of the generation of additional non-linearities,
which in manifold physics has been successfully handled through
analysis of KPZ type equations and hopefully may lead to new physics.

Finally this field opened at least two Pandora's boxes at the face of
the theorist. $d=2$ has prompted for the very question of the validity
of a naive renormalization group in a glassy system, when many
metastable states exist. This question and idea that RG should
in some way be complemented by the proper RSB procedure has
been investigated subsequently in several other disordered systems.
Although we still do not know how to mix these two
ingredients efficiently, simple models such as the CO
could be good laboratories
to try, with the advantage of being directly experimentally relevant.
Last but not least, the moving glass was the first hint that
non potential dynamics can indeed exhibit glassy properties,
a question one can ask in a more general context. The very
existence of such {\it dissipative glasses} may seem at first sight
unnatural because after all there is constant dissipation going on in the
system which naturally tends to generate or increase the effective
temperature and kill glassy
properties. There too the situation may be more subtle and leave
much room for unexpected behavior. The
study of the properties of such dissipative glasses and their
comparison with normal ones, will doubtless prove to be a
field worth tapping.


\end{document}